\documentstyle[epsf,12pt]{article}
\begin{document}
\begin{center}
\Large{\bf The Photon and Light $1^{--}$ Mesons.}\\
\large{R.S. Longacre$^a$\\
$^a$Brookhaven National Laboratory, Upton, NY 11973, USA}
\end{center}
 
\begin{abstract}
In this paper we look into the world data related to the photon and the 
connection to light mesons with the same spin parity quantum number as the
photon $J^p = 1^-$.
\end{abstract}
 
\section{Introduction to $\pi$ $\pi$ scattering cross section} 

In this paper we use STAR high precision Au + Au ultra-peripheral 
coherent photoproduction collisions at $\sqrt{s_{NN}} =$ 200 GeV(the highest 
RHIC energy)\cite{photo}. These photoproduction collisions produce pairs of
pions($\pi^+$ $\pi^-$ pairs) which are mainly in a $J^{pc} = 1^{--}$ quantum
number system. We perform a global fit to the photoproduction combined with
$e^+$ $e^-$ scattering to $\pi^+$ $\pi^-$\cite{eepipi} and 
$\pi^+$ $\pi^-$ $\pi^0$\cite{eepipipi}. To this fit we also use p-wave partial
wave analysis of $\pi^+$ $\pi^-$ to $\pi^+$ $\pi^-$\cite{pipipipi}.

The paper is organized in the following manner:

Sec. 1 is a introduction to $\pi$ $\pi$ scattering cross section. Sec. 2 
addresses re-scattering through the P-wave $\pi$ $\pi$ which is  a 
multi-channel problem. Sec. 3 is a global S$\rm\ddot o$ding model fit to all 
the Data. Sec. 4 we find 5 S-matrix poles in the $J^{pc} = 1^{--}$ quantum 
system. Sec. 5 we go beyond the S$\rm\ddot o$ding model and break up the 
photoproduction amplitude into its component parts and show how they are 
different in the lower mass and higher mass regions. Sec. 6 presents the 
summary and discussion.

\subsection{$\pi$ $\pi$ scattering cross section} 

For the first part of this story we will define what a scattering
cross section is. We will first only consider elastic
scattering of pions. Two pions can scatter at a certain energy
which we will call $M_{\pi \pi}$. The differential cross section
$\sigma$ at a given $M_{\pi \pi}$ is
\begin{equation}
\frac{d\sigma}{d\phi d\theta} = \frac{1}{K^2} \left| \sum_{\ell}
(2\ell + 1) T_{\ell} P_{\ell} (cos\theta) \right| ^2
\end{equation}
where $\phi$ and $\theta$ are the azimuthal and scattering angles,
respectively. $T_{\ell}$ is a complex scattering amplitude and
$\ell$ is the angular momentum. $P_{\ell}$ is the Legendre
polynomial, which is a function of $cos\theta$. $K$ is the flux
factor equal to the pion momentum in the center of mass. The $T_{\ell}$
elastic scattering amplitudes are complex amplitudes
described by two real numbers one bounded between 0.0 and 1.0($\eta_{\ell}$) 
and another is in units of angles($\delta_{\ell}$). The form of the amplitude is
\begin{equation}
T_{\ell} = \eta_{\ell} sin\delta_{\ell}e^{i \delta_{\ell}}
\end{equation}
We note that $\eta_{\ell}$ and $\delta_{\ell}$ depends on the value of $\ell$ 
and $M_{\pi\pi}$. 

\section{Re-scattering through the P-wave $\pi$ $\pi$ is a multi-channel 
problem}

In order to handle the multi-channel problem which exist in P-wave 
$J^{pc} = 1^{--}$ we will use the K-matrix approach. When we are below the 
$\pi$  $\pi$ $\pi$  threshold the system is only a one channel problem and the 
K-matrix is only a single term of the matrix however as we move up in energy 
more channels open up. Also the $1^{--}$ couples to the photon which can couple
to pairs of electrons($e^+$$e^-$). and muons($\mu^+$$\mu^-$). Below is a table
of channels we will use in our K-matrix approach.

\bf Table I. \rm The channels used and the number assigned to them.

\begin{center}
\begin{tabular}{|c|r|}\hline
\multicolumn{2}{|c|}{Table I}\\ \hline
channel particles & channel number \\ \hline
$\pi^+$$\pi^-$ & 1 \\ \hline
$\pi^+$$\pi^-$$\pi^0$ & 2 \\ \hline
$e^+$$e^-$ & 3 \\ \hline
$\mu^+$$\mu^-$ & 4 \\ \hline
$\eta$ $\pi^+$$\pi^-$ & 5 \\ \hline
\end{tabular}
\end{center}

Ordinarily in a hadronic K-matrix there would be a unique set of quantum 
numbers which includes isospin. However with the coupling of the photon we 
have isospin mixing. The isospin of the $\pi^+$ $\pi^-$ system is $I = 1$. 
The $\pi^+$ $\pi^-$ $\pi^0$ system is $I = 0$. The $\eta$ $\pi^+$ $\pi^-$ 
system is $I = 1$. The $e^+$$e^-$ and $\mu^+$$\mu^-$ couples to both isospins
and have a universal coupling through the photon. 

The $1^{--}$ P-wave $\pi$ $\pi$ couples mainly to the $\rho(760)$ and can with
first blush be considered a single channel problem. Such a k-matrix would be 
given by,

\begin{equation}
K_{11} = tan\delta_0.
\end{equation}

We see that when $\delta_0$ = $90^\circ$ that there will be a pole in the
K-matrix and in this case the $\rho(760)$. 

\begin{equation}
K_{11} = {\frac{2\gamma_1^2Z_1^2q_1}{(Z_1^2+1)M_{\pi \pi}}\over{(M_1^2 - M_{\pi \pi}^2)}}.
\end{equation}

Where $\gamma_1$ is the coupling of the pole($\rho(760)$)  to the $\pi \pi$ 
channel, $q_1$ = $q_{\pi \pi}$ is the center of mass momentum of the $\pi \pi$ 
channel with $Z_1$ as the ratio of the center mass momentum $q_1$
divided by $q_s$ which is .200 GeV/c(size of 1.0 fm),  also $M_1$ is the 
mass of the pole($\rho(760)$), The T-matrix is given by

\begin{equation}
T_{11} = e^{i \delta_0} sin\delta_0 = {K_{11}\over{(1 - iK_{11})}} = \,(1 - iK)^{-1}K.
\end{equation}

\subsection{Low Mass Region(threshold to 1.1Gev)} 

The $\rho(760)$ couples to four of the five channels which we consider in this
paper. Each of channels have a phase space factor. For channel 1($\pi \pi$)
we had $\frac{2Z_1^2q_1}{(Z_1^2+1)M_{\pi \pi}}$ on the previous page. For 
channel 2($\pi \pi \pi$) we have $\frac{2Z_2^2q_2}{(Z_2^2+1)M_{\pi \pi}}$, where
$q_2$ is the center mass momentum of two pions at rest to the other pion. Here 
we use $m_{\pi \pi}$ to register energy of the system which is the same in all 
channels. $Z_2$ is the ratio of $q_2$ to $q_s$. Channel 3($e^+ e^-$) we have 
$\frac{2q_3}{M_{\pi \pi}}$, where $q_3$ is the center mass momentum of the 
$e^+ e^-$ system. Channel 4($\mu^+ \mu^-$) we have $\frac{2q_4}{M_{\pi \pi}}$, 
where $q_4$ is the center mass momentum of the $\mu^+ \mu^-$ system.

Isospin mixing is part of this four channel $j^{pc}$ = $1^{--}$. the isospin
partner of the $\rho(760)$ is the $\omega(783)$. We need two k-matrix poles
where pole 1 is the $\rho(760)$ and pole 2 is the $\omega(783)$. The k-matrix
is given by

\begin{equation}
K_{11} = \sum_{i} {\frac{2\gamma_{i1}^2Z_1^2q_1}{(Z_1^2+1)M_1}\over{(M_i^2 - M_1^2)}},
K_{22} = \sum_{i} {\frac{2\gamma_{i2}^2Z_2^2q_2}{(Z_2^2+1)M_1}\over{(M_i^2 - M_1^2)}},
K_{33} = \sum_{i} {\frac{2\gamma_{i3}^2q_3}{M_1}\over{(M_i^2 - M_1^2)}},
K_{44} = \sum_{i} {\frac{2\gamma_{i4}^2q_4}{M_1}\over{(M_i^2 - M_1^2)}}.
\end{equation}
 
An example of a cross term

\begin{equation}
K_{12} = \sum_{i} {\frac{2\gamma_{i1}\gamma_{i2}\sqrt{Z_1^2Z_2^2q_1q_2}}{\sqrt{(Z_1^2+1)(Z_2^2+1)M_1^2}}\over{(M_i^2 - M_1^2)}}.
\end{equation}

The T-matrix is given by

\begin{equation}
T  = \,(\delta - iK)^{-1}K.
\end{equation}

The universal lepton coupling through the photon means that $\gamma_{i3}$ = 
 $\gamma_{i4}$. 

\subsection{High Mass Region(1.1Gev to 1.9Gev)} 

For the high mass region we use a factorizable 
k-matrix($k_1$,$k_2$,$k_3$$k_4$,$k_5$), such that

\begin{equation}
K_{ij} = k_i k_j.
\end{equation}

We use a $9^{th}$ order polynomial as a function of $M_{\pi \pi}$ for $k_1$ and
staring with the values generated from Ref.\cite{pipipipi} see the appendix.
For $k_2$ we use zero coupling assuming that the only three pion state is the 
$\omega(783)$ which is very narrow and confined to the lower mass region.
for $k_3$ and $k_4$ we use the same simple linear function of $M_{\pi \pi}$ 
since we have universal lepton coupling through the photon. For $k_5$ we
use the $9^{th}$ order polynomial staring values generated from 
Ref.\cite{pipipipi} see the appendix where we use for $k_5$ the k-matrix 
element denoted by $k_2$.

\begin{figure}
\begin{center}
\mbox{
   \epsfysize 8.0in
   \epsfbox{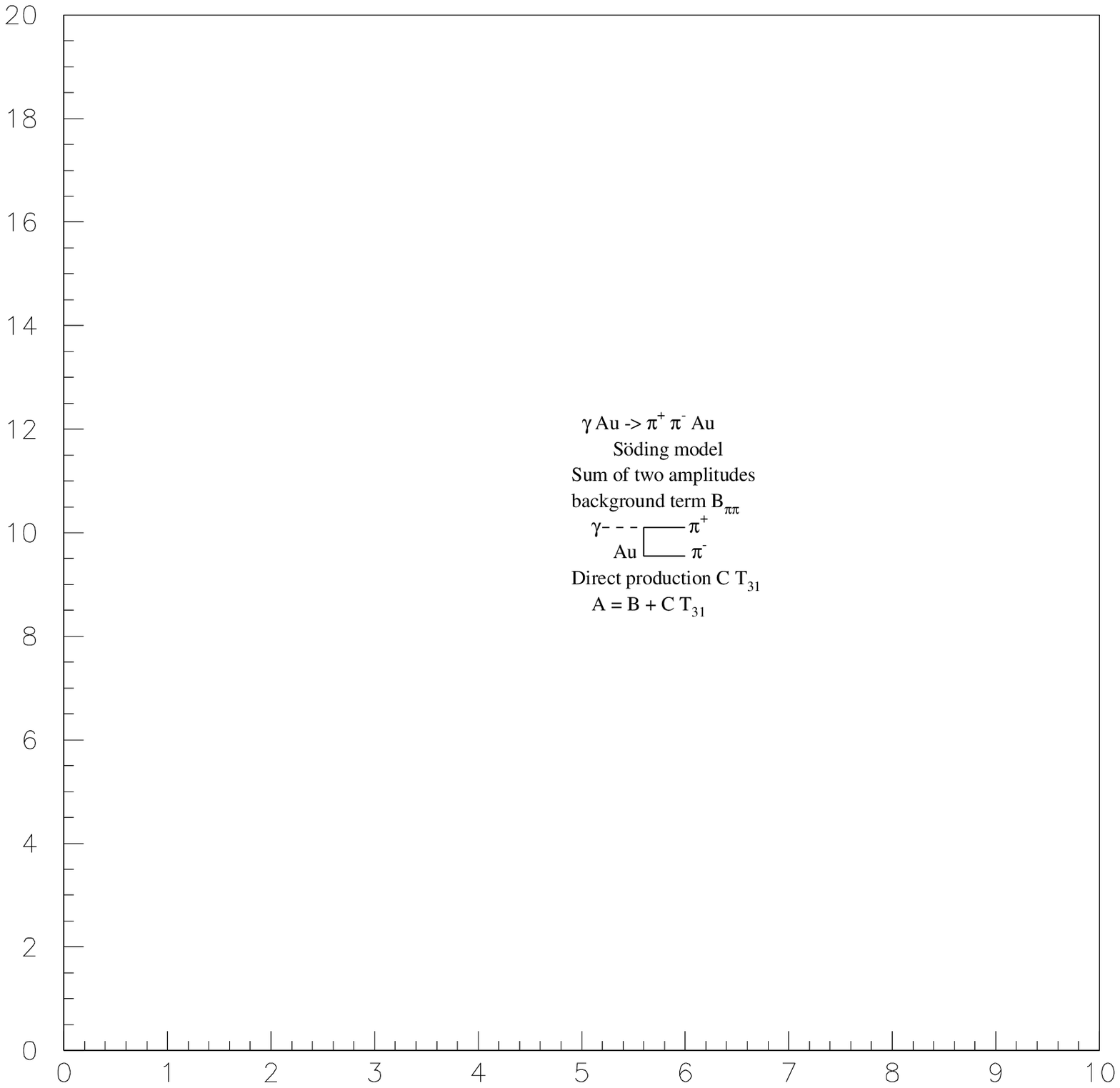}}
\end{center}
\vspace{2pt}
\caption{The S$\rm\ddot o$ding Model\cite{soding} is made up of two terms.
the first term is a direct formation of $\pi^+$ $\pi^-$ from the photon, which
is depicted in the diagram above. The second term is the photon going into
quarks and then into $\pi^+$ $\pi^-$. For this term we use $e^+$ $e^-$ 
scattering to $\pi^+$ $\pi^-$\cite{eepipi} as stand in for this 
formation($T_{31}$).}
\label{fig1}
\end{figure}

\begin{figure}
\begin{center}
\mbox{
   \epsfysize 8.0in
   \epsfbox{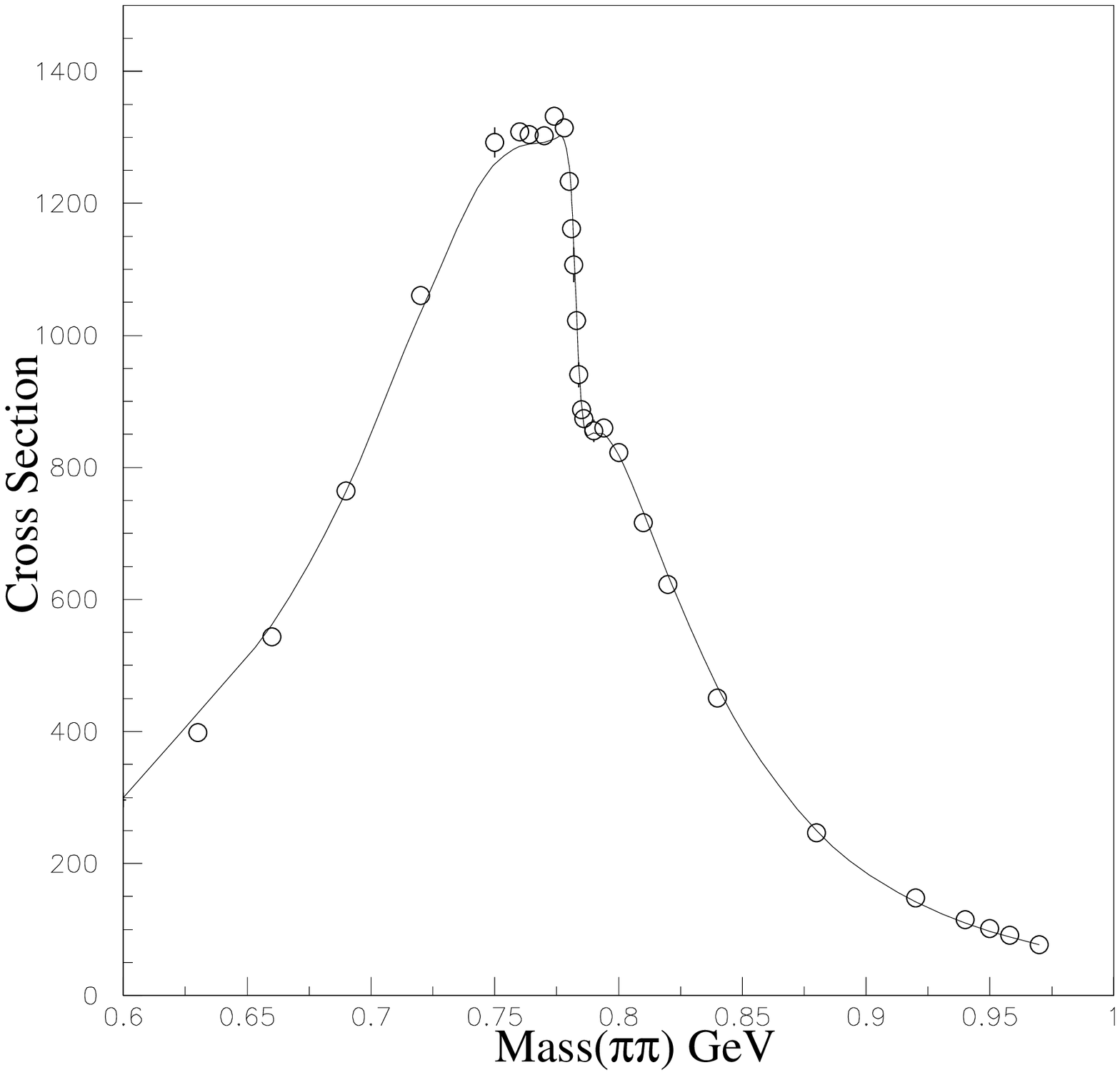}}
\end{center}
\vspace{2pt}
\caption{The scattering cross section of $e^+$ $e^-$ scattering to $\pi^+$ 
$\pi^-$\cite{eepipi} and our global fit described in the text.}
\label{fig2}
\end{figure}

\begin{figure}
\begin{center}
\mbox{
   \epsfysize 8.0in
   \epsfbox{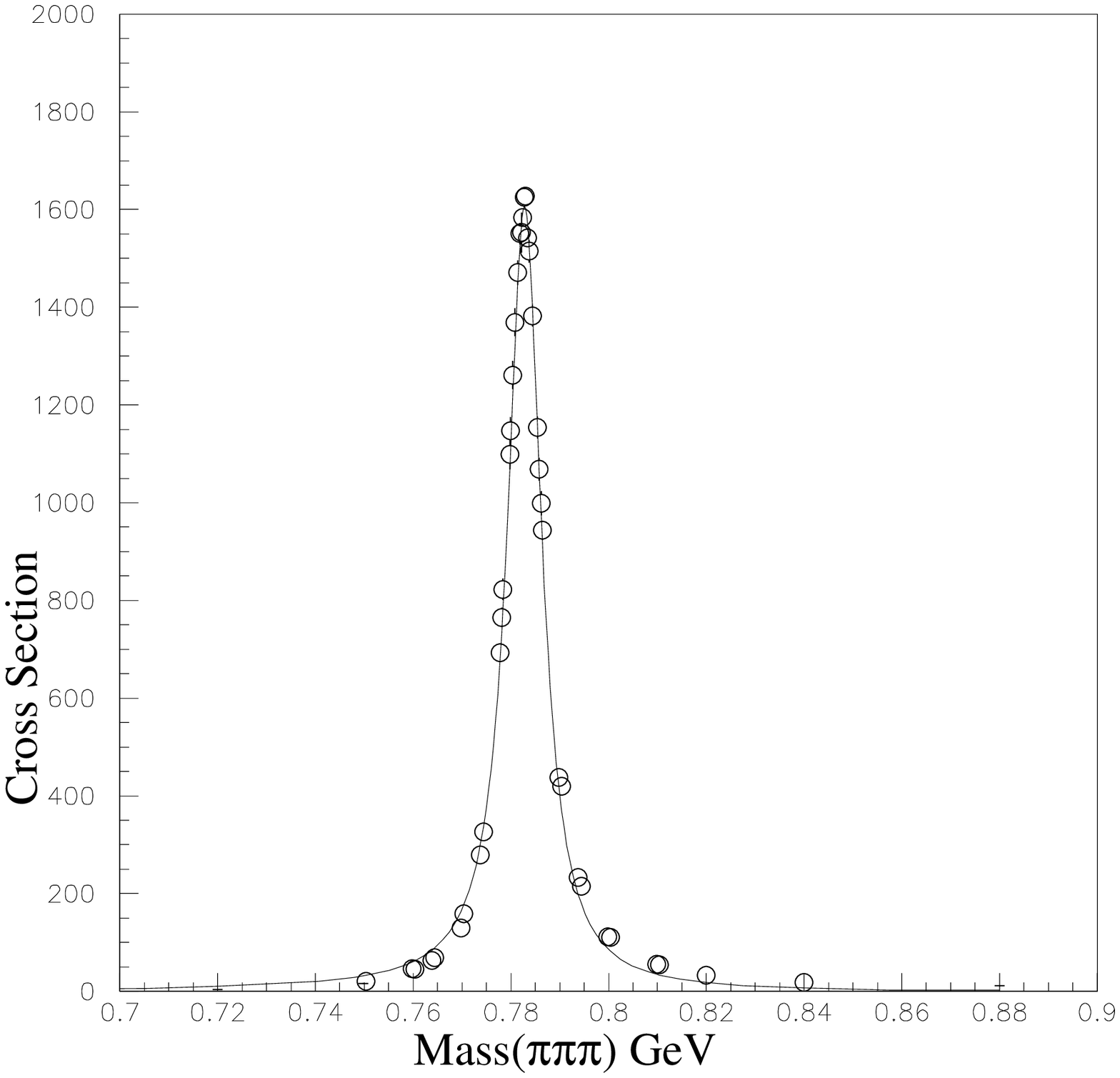}}
\end{center}
\vspace{2pt}
\caption{The scattering cross section of $e^+$ $e^-$ scattering to 
$\pi^+$ $\pi^-$ $\pi^0$\cite{eepipipi} and our global fit described in the 
text.}
\label{fig3}
\end{figure}

\section{A Global S$\rm\ddot o$ding Model Fit to all the Data}

We perform a global fit to the coherent photoproduction data using a 
generalized S$\rm\ddot o$ding Model\cite{soding} plus $e^+$ $e^-$ scattering 
to $\pi^+$ $\pi^-$\cite{eepipi} and $\pi^+$ $\pi^-$ $\pi^0$\cite{eepipipi}. 
To this fit we also use p-wave partial wave analysis of $\pi^+$ $\pi^-$ to 
$\pi^+$ $\pi^-$\cite{pipipipi}. The 4 channel k-matrix is used for low mass 
region and the 5 channel for high mass region as set up in last section. 
We fit directly to the $T_{11}$ Argand plot of Ref.\cite{pipipipi}. As part of 
the global fit we also fit Ref.\cite{eepipi} and Ref.\cite{eepipipi} using

\begin{equation}
Cross Section_{\pi \pi} = {{C}\over{p_{ee}}} \vert T_{31} \vert^2,
\end{equation}

and

\begin{equation}
Cross Section_{\pi \pi \pi} = {{C}\over{p_{ee}}} \vert T_{32} \vert^2.
\end{equation}

The photoproduction data are fitted using a generalized S$\rm\ddot o$ding 
Model\cite{soding} which is outlined in Figure 1. We see in the figure that
there are two terms. One being a S$\rm\ddot o$ding background term of pion
pairs coming directly out of the vacuum plus a second being the photon 
coupling directly to the $1^{--}$ scattering matrix. The universal lepton pair
coupling through the photon gives us $T_{31}$ as the direct coupling.
We use the form for the S$\rm\ddot o$ding background term given by

\begin{equation}
B_{\pi\pi} = b_{\pi\pi}\sqrt{{2p_{\pi\pi}}\over{M_{\pi\pi}}} e^{-d_{\pi\pi}M_{\pi\pi}}.
\end{equation}

The directly coupled term of the photon to the S-matrix is given by

\begin{equation}
C T_{31} = c\sqrt{{M_{\pi \pi}}\over{2p_{e e}}} T_{31}.
\end{equation}

The results of the global fit is shown in Figures 2 through 6. The overall 
$\chi^2$ for the global fit which has 1669 degrees of freedom is 1926. The 
1 $\sigma$ error on such a fit is a $\Delta \chi^2$ of 58. This implies 
that this global fit is 4.8$\sigma$ away from a 1669 $\chi^2$ fit.

\begin{figure}
\begin{center}
\mbox{
   \epsfysize 8.0in
   \epsfbox{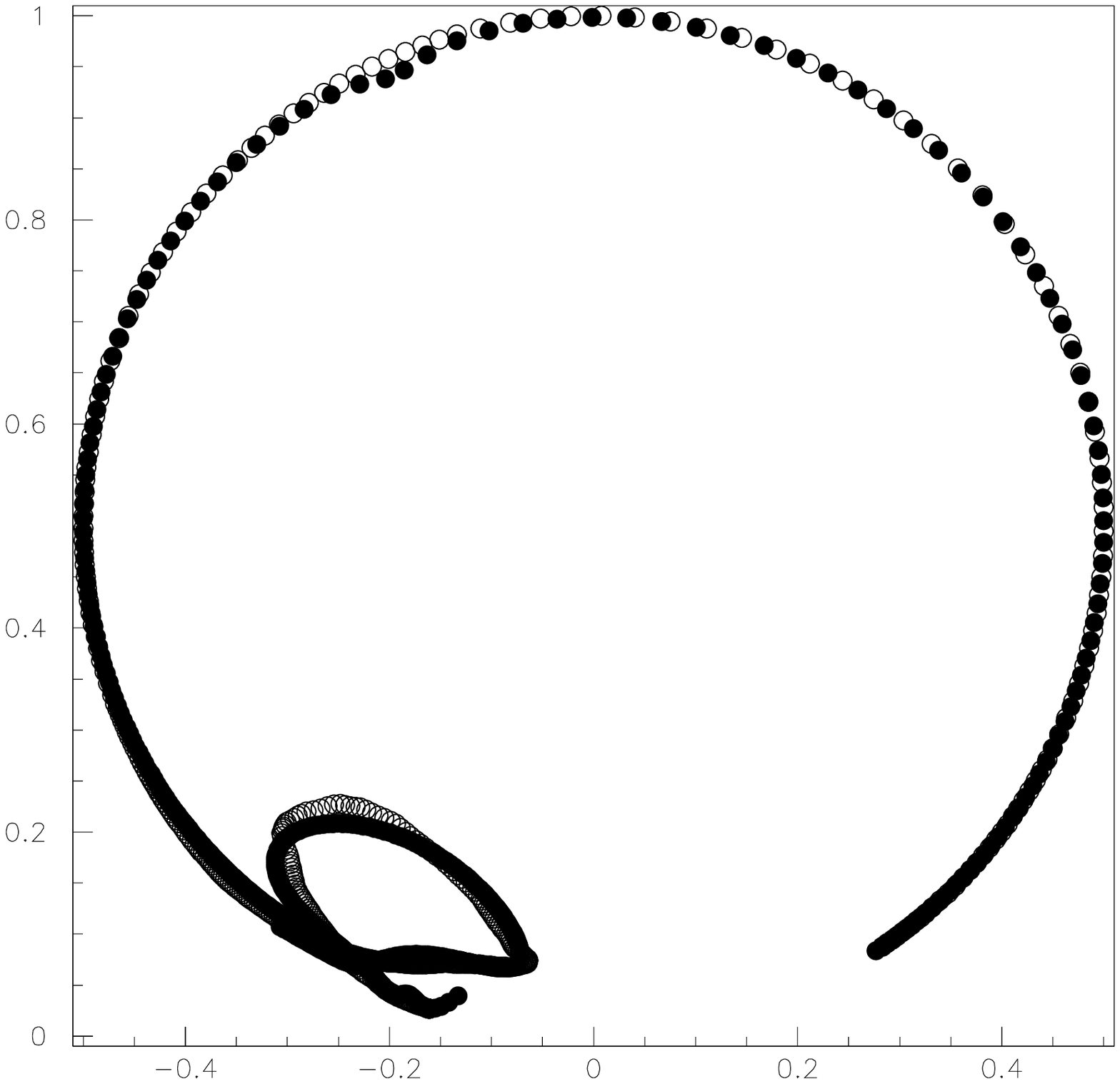}}
\end{center}
\vspace{2pt}
\caption{The Argand plot of the S-matrix p-wave partial wave analysis of 
$\pi^+$ $\pi^-$ to $\pi^+$ $\pi^-$\cite{pipipipi}. The input amplitude
are open dots($\circ$) and our global fit described in the text are solid dots 
($\bullet$).}
\label{fig4}
\end{figure}

\begin{figure}
\begin{center}
\mbox{
   \epsfysize 8.0in
   \epsfbox{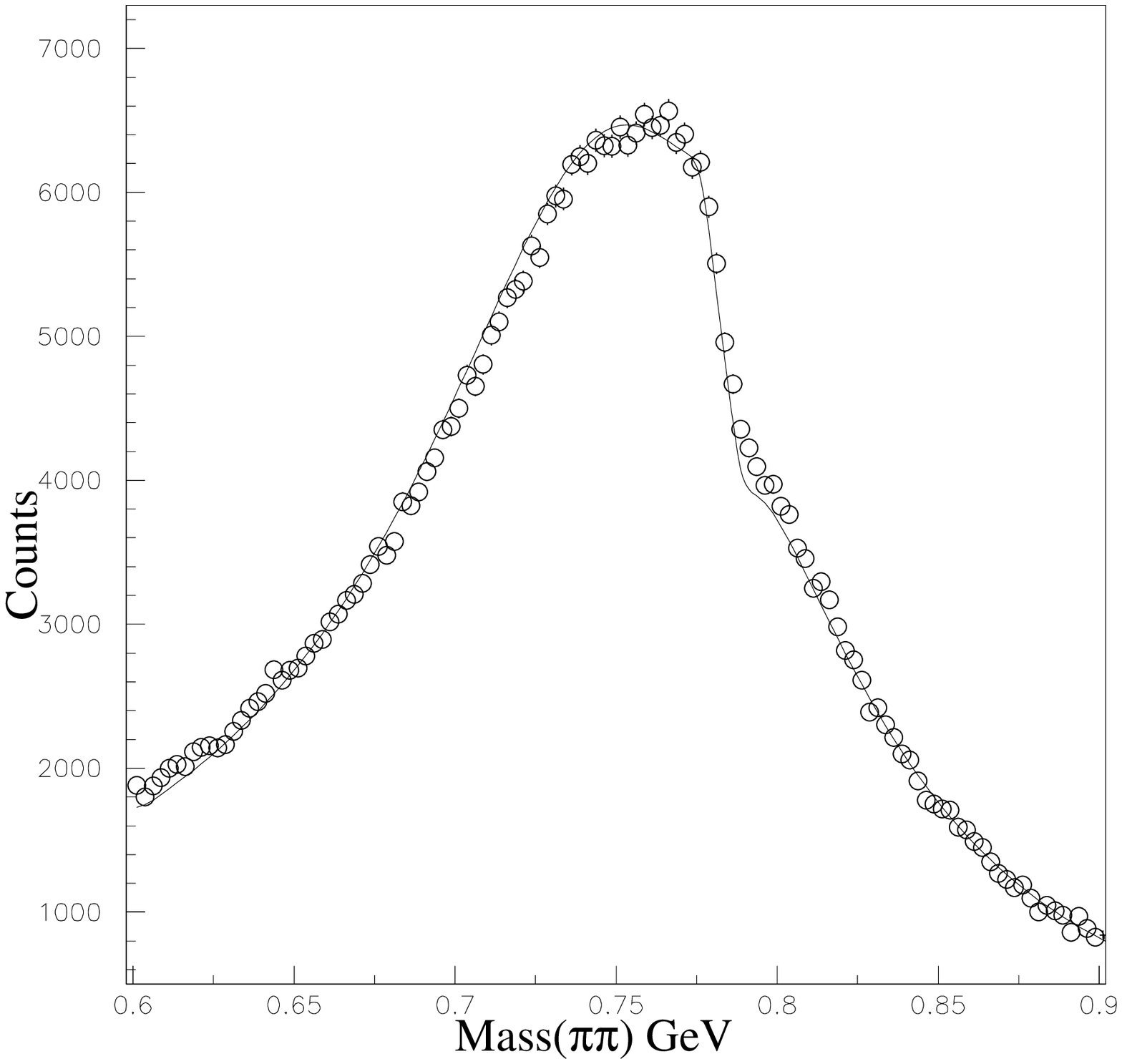}}
\end{center}
\vspace{2pt}
\caption{ The STAR high precision Au + Au ultra-peripheral coherent 
photoproduction data at $\sqrt{s_{NN}} =$ 200 GeV(the highest RHIC energy) for 
the lower mass region 0.6 GeV to 0.9 GeV and our global fit described in the 
text.}
\label{fig5}
\end{figure}

\begin{figure}
\begin{center}
\mbox{
   \epsfysize 8.0in
   \epsfbox{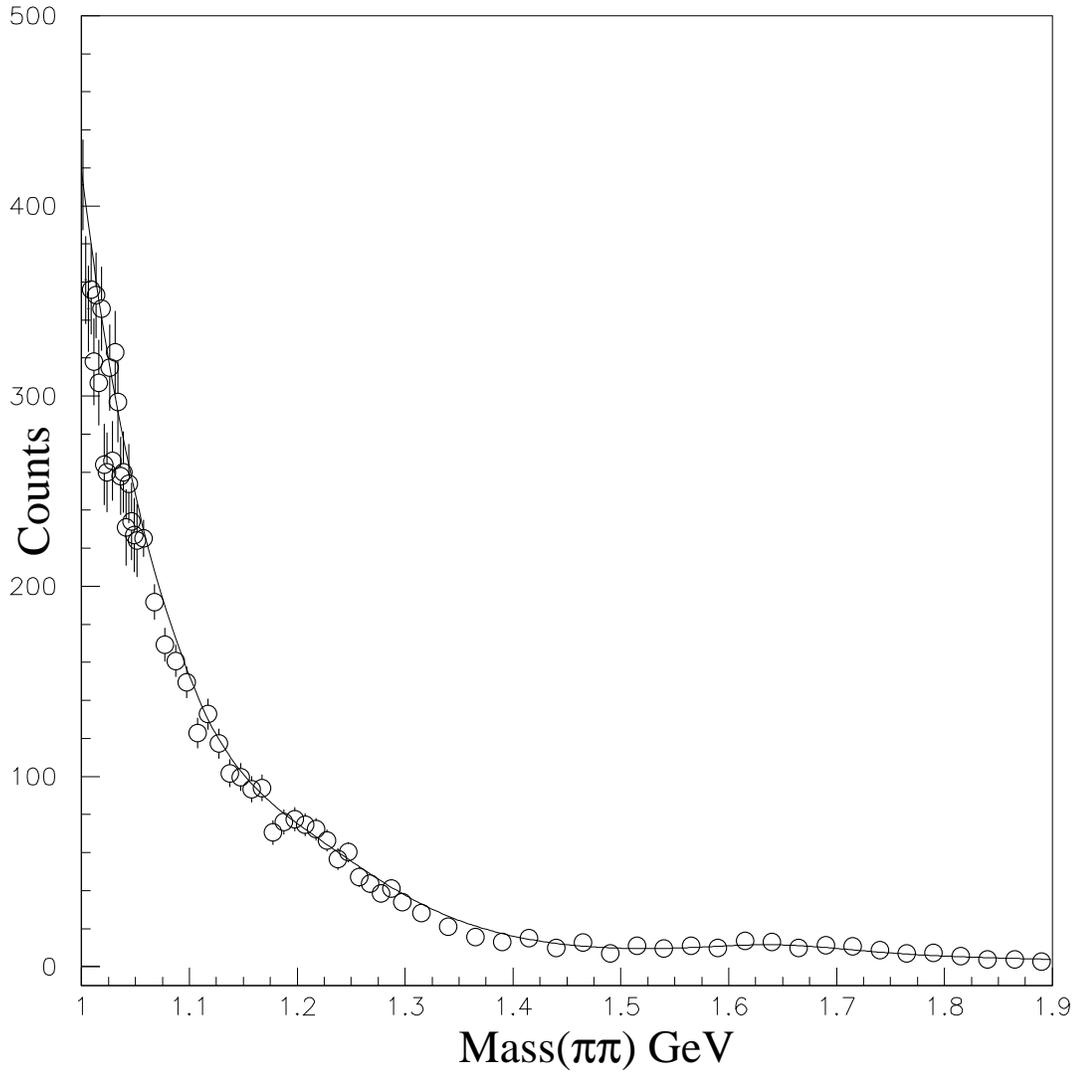}}
\end{center}
\vspace{2pt}
\caption{ The STAR high precision Au + Au ultra-peripheral coherent
photoproduction data at $\sqrt{s_{NN}} =$ 200 GeV(the highest RHIC energy) for 
the higher mass region 1.0 GeV to 1.9 GeV and our global fit described in the 
text.}
\label{fig6}
\end{figure}

\clearpage

\section{Poles of the $1^{--}$ S-Matrix}

In the above section we have achieve a global fit to the data sets which 
includes the STAR high precision Au + Au ultra-peripheral coherent  
photoproduction collisions into pairs of pions($\pi^+$ $\pi^-$ pairs) plus
$e^+$ $e^-$ scattering to $\pi^+$ $\pi^-$\cite{eepipi} and 
$\pi^+$ $\pi^-$ $\pi^0$\cite{eepipipi}. To this fit we also fit p-wave partial
wave analysis of $\pi^+$ $\pi^-$ to $\pi^+$ $\pi^-$\cite{pipipipi}. With this
fit we have achieve an analytic form for the S-matrix. Using this analytic 
form we search for poles in the $1^{--}$ S-matrix. In this search we find 5
poles. 

\bf Table II. \rm Poles of the $1^{--}$ S-Matrix. 

\begin{center}
\begin{tabular}{|c|r|r||r|r|r|}\hline
\multicolumn{6}{|c|}{Table II}\\ \hline
mass & width & $\pi$ $\pi$ & $\pi$ $\pi$ $\pi$ &  $e^+$ $e^-$ & other \\ \hline
$0.764 GeV$ & $.148 GeV$ & 100 & 0 & .0047 & 0 \\ \hline
$0.783 GeV$ & $.008 GeV$ & 2 & 89 & .0074 & 9 \\ \hline
$1.333 GeV$ & $.334 GeV$ & 5.4 & 0 & .0000015 & 94.6 \\ \hline
$1.629 GeV$ & $.221 GeV$ & 19 & 0 & .0000039 & 81 \\ \hline
$1.733 GeV$ & $.246 GeV$ & 7.5 & 0 & .00000083 & 92.5 \\ \hline
\end{tabular}
\end{center}

The first pole is the well know $\rho (760)$, while its isospin partner
$\omega (780)$ is the second pole. In the table for the $\omega (780)$ the 
decay mode for this state marked as other is the radiative decay mode 
$\gamma$ $\pi$. The branching factions are measured in percentage.

In the high mass region one expects there should be two radial excitations of 
the $\rho (760)$. These excitation masses should be at around 1300 and 1800 MeV.
One should also expect there should be a d-wave $q$ $\bar q$ state at around 
1600 MeV. We see the poles in the high mass are consistent with this picture.
.
\begin{figure}
\begin{center}
\mbox{
   \epsfysize 8.0in
   \epsfbox{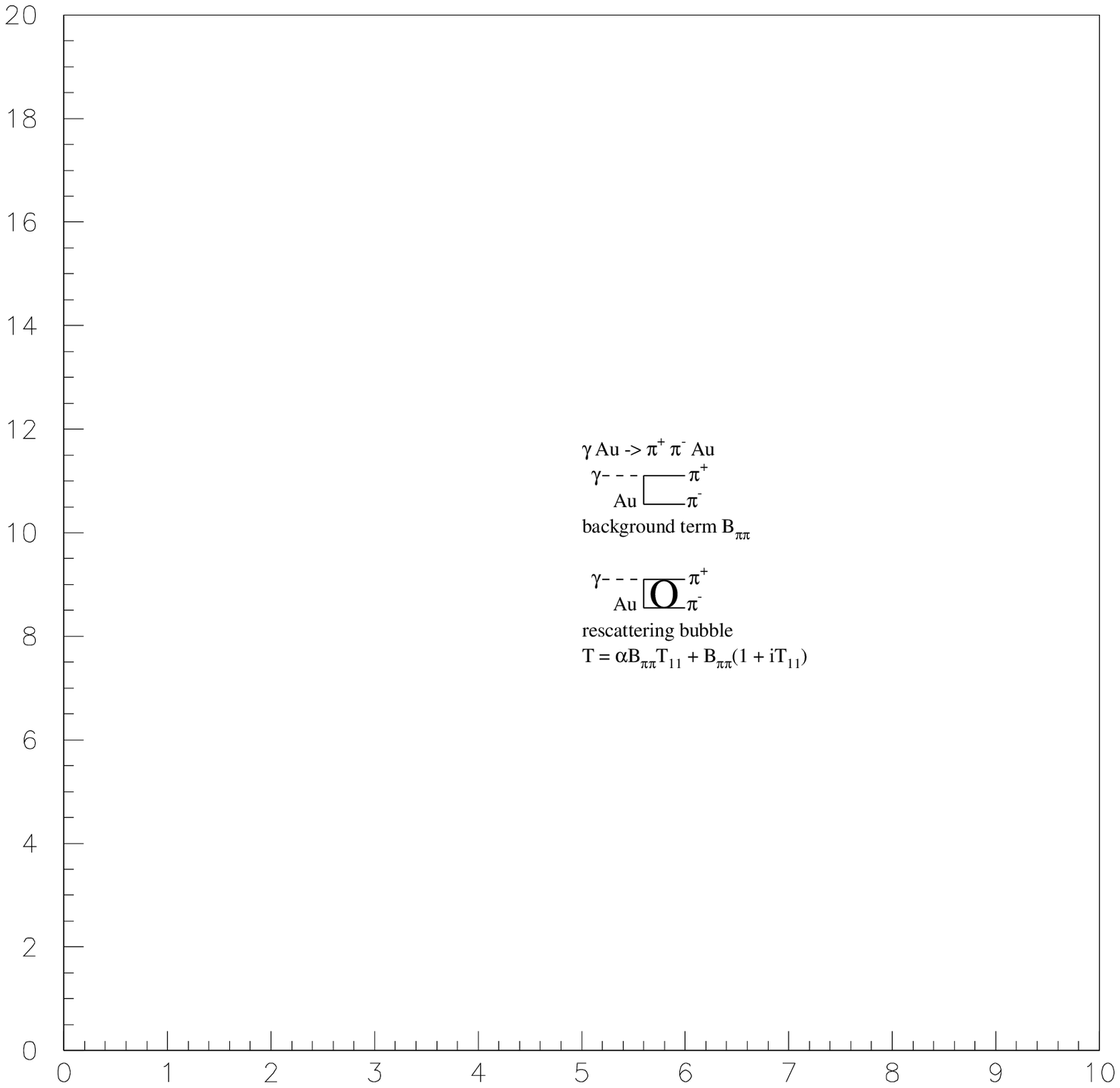}}
\end{center}
\vspace{2pt}
\caption{Beyond the S$\rm\ddot o$ding Model we focus on two terms one being a 
direct formation of $\pi^+$ $\pi^-$ from the photon and the other being
the re-scattering through the S-matrix of $\pi$ $ \pi$ p-wave scattering. 
The loop or bubble has a real and a imaginary part. This leads to the above 
equation.}
\label{fig7}
\end{figure}

\section{Beyond the S$\rm\ddot o$ding Model Fit.}

The S$\rm\ddot o$ding Model has a built-in term of interference between the
direct production term and the background term. This interference term is
equivalent to a re-scattering as shown in Figure 7. The re-scattering term 
is a loop of pions coming from the background and re-interacting with the
S-matrix of $\pi$ $ \pi$ p-wave scattering. The loop or bubble has a real and a 
imaginary part. The imaginary part of the loop is equal to 
$B_{\pi\pi}$(equation 12) times S-matrix $\pi$ $ \pi$ p-wave scattering. The 
real part of the loop is equal to $\alpha B_{\pi\pi}$ times S-matrix $\pi$ $\pi$
scattering. The value of $\alpha$ is determined to be 2.0 in Ref.\cite{loop} 
from photo production data. Figure 7 and equation 14 gives the equation for 
the background plus re-scattering 

\begin{equation}
T = 2.0 B_{\pi\pi} T_{11} + B_{\pi\pi} (1.0 + i T_{11}).
\end{equation}

\clearpage

\subsection{The Lower Mass region}

To the above term we must add the direct production of the $\rho (760)$
pole and the $\omega (780)$ pole in the lower mass region. When this is done
we break the constraint of the S$\rm\ddot o$ding Model that the 
$\omega (780)$ into $\pi$ $\pi$ has to be the same as $e^+$ $e^-$ 
scattering to $\pi^+$ $\pi^-$\cite{eepipi}. With this added freedom the 
global fit $\chi^2$ improve by 257 which a 4.8 $\sigma$ 
improvement(see Figure 8). This photo production amplitude into 
$\pi$ $\pi$ is shown in Figure 9, while the S$\rm\ddot o$ding Model 
$\pi$ $\pi$ amplitude is shown in Figure 10. Even though we have a 
quantitative difference these amplitudes are qualitatively the same.

\subsection{The Higher Mass region}

In the higher mass region a new background becomes possible. In our model for
the $1^{--}$ system we have a large coupling of the $\pi$ $\pi$ channel to the
$\eta$ $\pi$ $\pi$ channel. Thus if the photon directly produces the $\eta$ 
$\pi$ $\pi$ channel, then through a $\eta$ $\pi$ $\pi$ loop one can form
the $\pi$ $\pi$ channel by the cross term of the S-matrix of $\eta$ $\pi$ $ 
\pi$ to $\pi$ $ \pi$(see Figure 11). Again the loop or bubble has a real and 
a imaginary part. The imaginary part of the loop is equal to 
$B_{\eta\pi\pi}$(equation 15) times T-matrix $\eta$ $\pi$ $\pi$ to $\pi$ $\pi$
scattering. The real part of the loop is equal to 2.0 $B_{\eta\pi\pi}$ times 
T-matrix $\eta$ $\pi$ $\pi$ to $\pi$ $\pi$. Figure 11 and equation 16 gives 
the equation for the re-scattering  
  
\begin{equation}
B_{\eta\pi\pi} = b_{\eta\pi\pi}\sqrt{{2p_{\eta\pi\pi}}\over{M_{\pi\pi}}} e^{-d_{\eta\pi\pi}M_{\pi\pi}}.
\end{equation}

\begin{equation}
T = 2.0 B_{\eta\pi\pi} T_{51} + i B_{\eta\pi\pi} T_{51}.
\end{equation}

We perform a new global fit to the photoproduction data plus $e^+$ $e^-$ 
scattering to $\pi^+$ $\pi^-$\cite{eepipi} and $\pi^+$ $\pi^-$ 
$\pi^0$\cite{eepipipi} plus the p-wave partial wave analysis of 
$\pi^+$ $\pi^-$ to $\pi^+$ $\pi^-$. As part of this fit we use equation 14 and 
equation 16. Since the photoproduction data is the square of the amplitude 
there is an over all phase that is not determined. Let us choose equation 14
to determine this phase. The result of this fit gives a result for equation 14 
which is displayed in Figure 12. The background rises quickly at threshold and
then falls off with mass. The higher mass region is the focus of Figure 13.
We see that the background term equation 14 for the pion loop is mainly a
real function. Equation 16 which is the $\eta$ $\pi$ $\pi$ loop is also mainly
real and of the same magnitude at these higher masses. This background 
also rises quickly at threshold and then falls off with mass tailing to 
zero(see Figure 14).

\begin{figure}
\begin{center}
\mbox{
   \epsfysize 8.0in
   \epsfbox{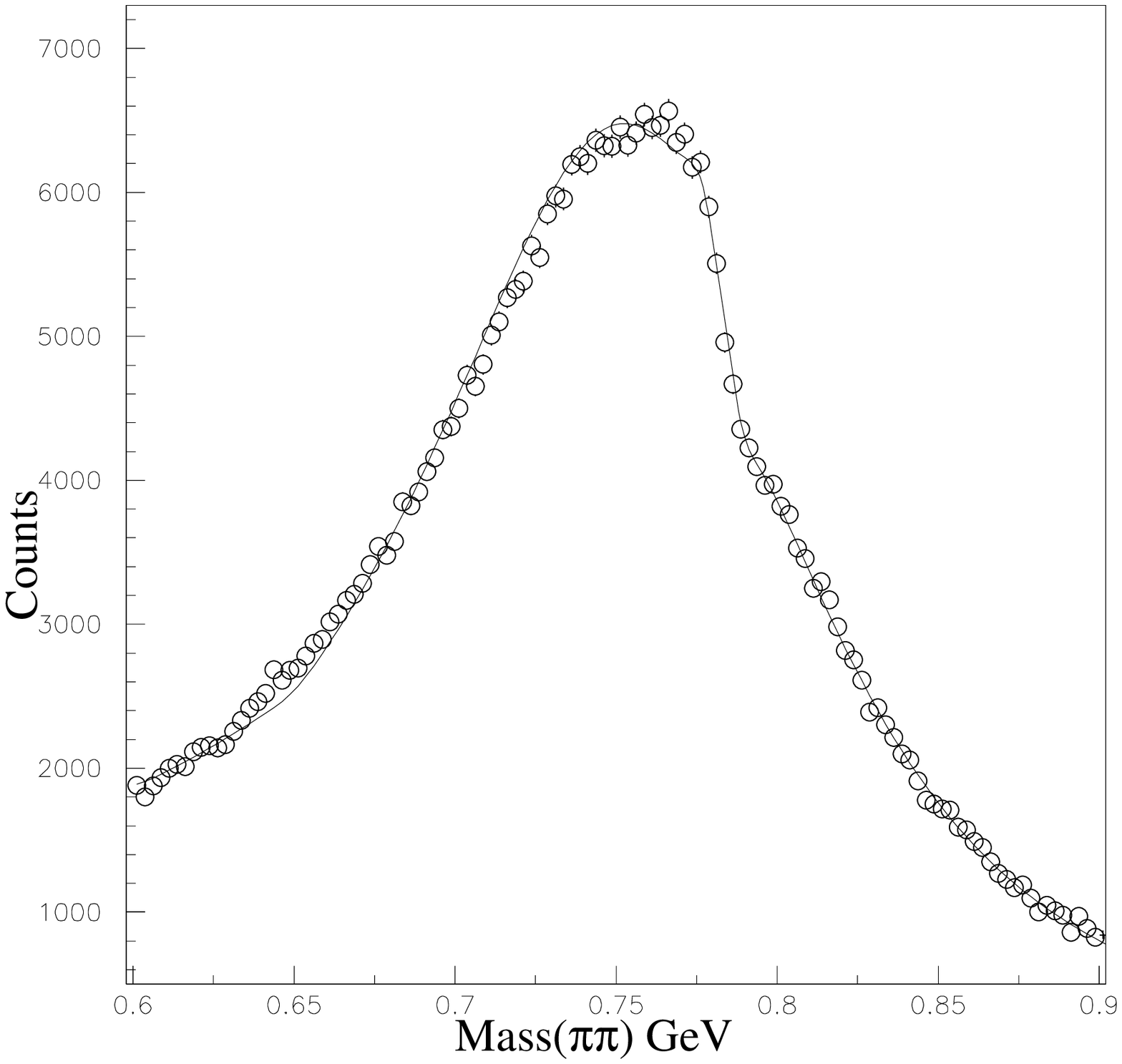}}
\end{center}
\vspace{2pt}
\caption{ The STAR high precision Au + Au ultra-peripheral coherent 
photoproduction data at $\sqrt{s_{NN}} =$ 200 GeV(the highest RHIC energy) for 
the lower mass region 0.6 GeV to 0.9 GeV and our global fit with re-scattering 
bubbles and direct production of poles as described in the text.}
\label{fig8}
\end{figure}

\begin{figure}
\begin{center}
\mbox{
   \epsfysize 8.0in
   \epsfbox{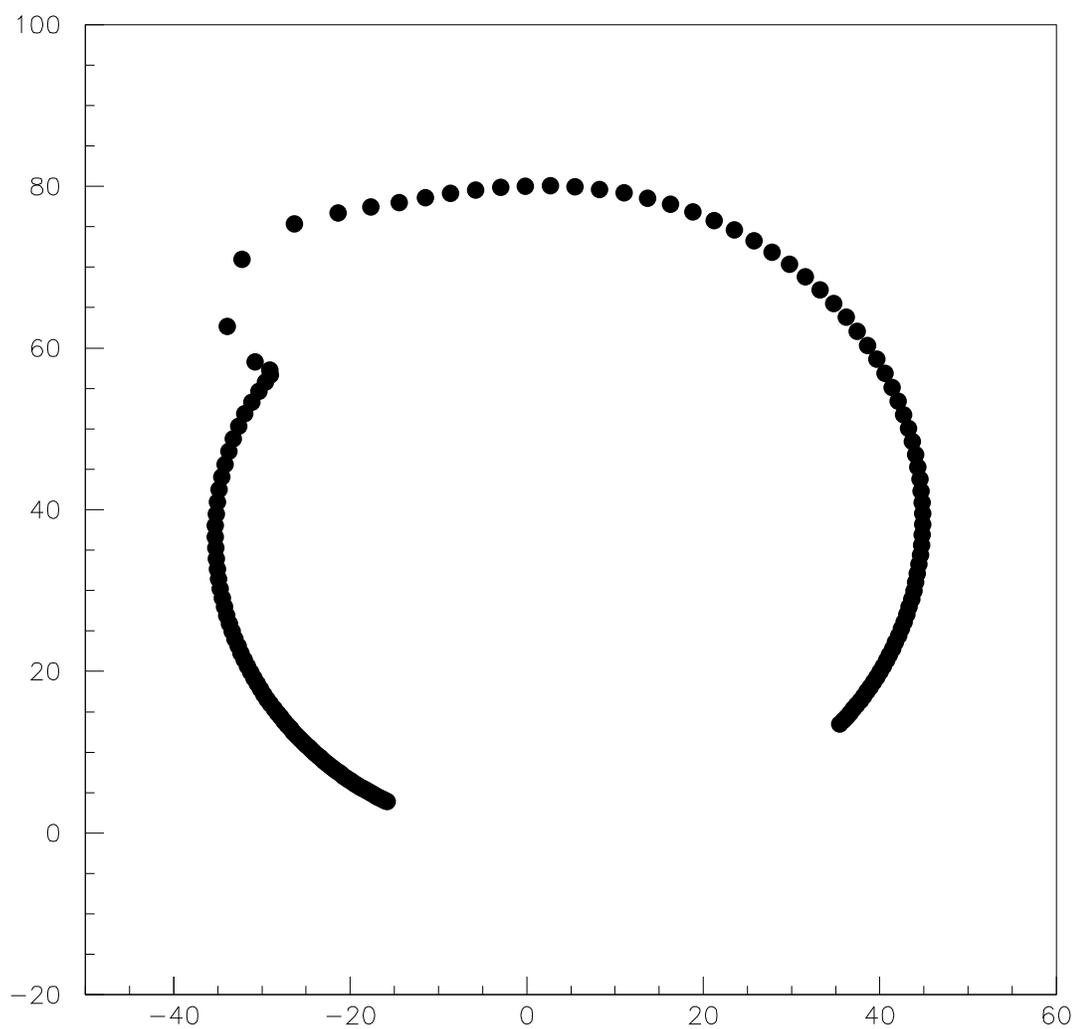}}
\end{center}
\vspace{2pt}
\caption{ The Argand plot of the amplitude of the photoproduction fit
with re-scattering bubbles and direct production of poles in the lower mass
region 0.6 GeV to 0.9 GeV as described in the text is shown.}
\label{fig9}
\end{figure}

\begin{figure}
\begin{center}
\mbox{
   \epsfysize 8.0in
   \epsfbox{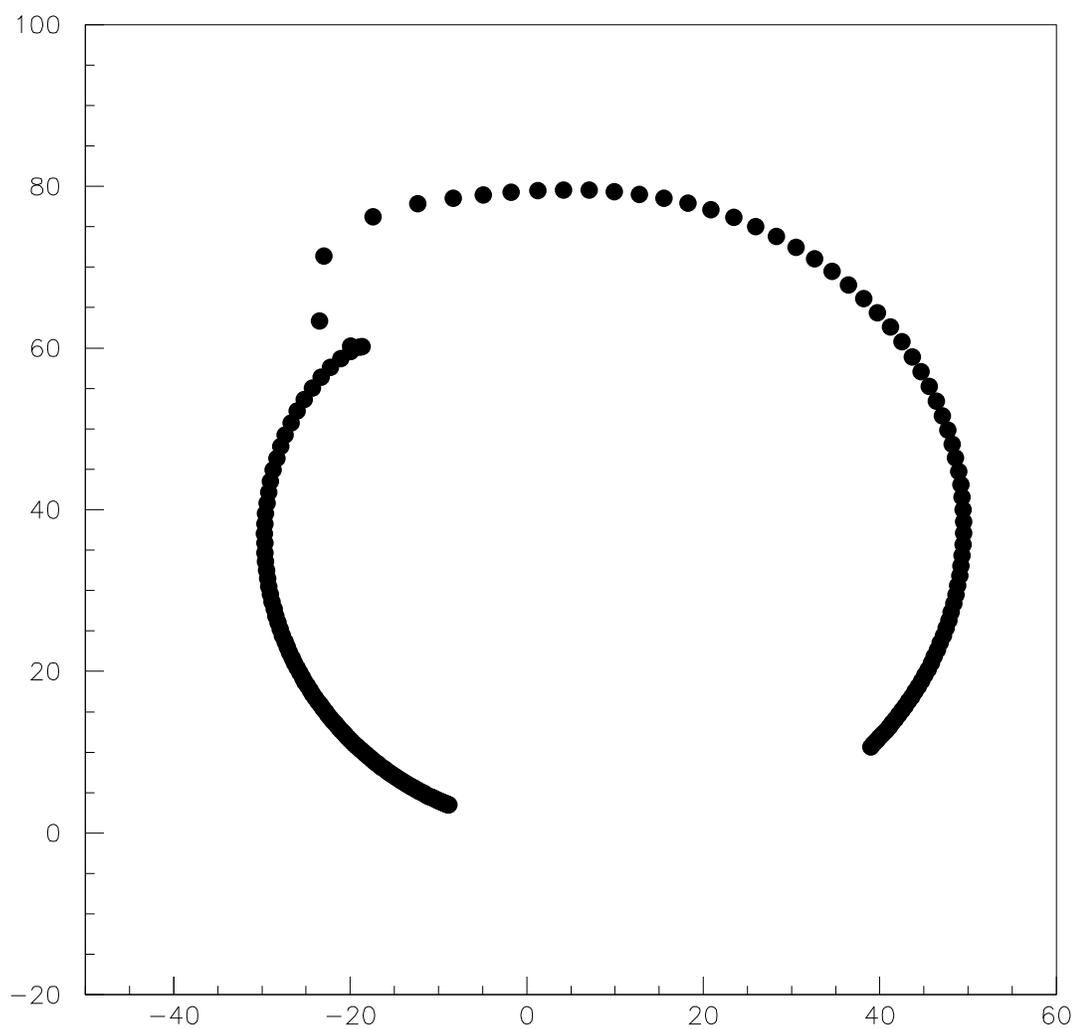}}
\end{center}
\vspace{2pt}
\caption{ The Argand plot of the amplitude of the photoproduction fit
from the S$\rm\ddot o$ding Model in the lower mass region 0.6 GeV to 0.9 GeV 
as described in the text is shown and should be compared with Figure 9.}

\label{fig10}
\end{figure}

\begin{figure}
\begin{center}
\mbox{
   \epsfysize 8.0in
   \epsfbox{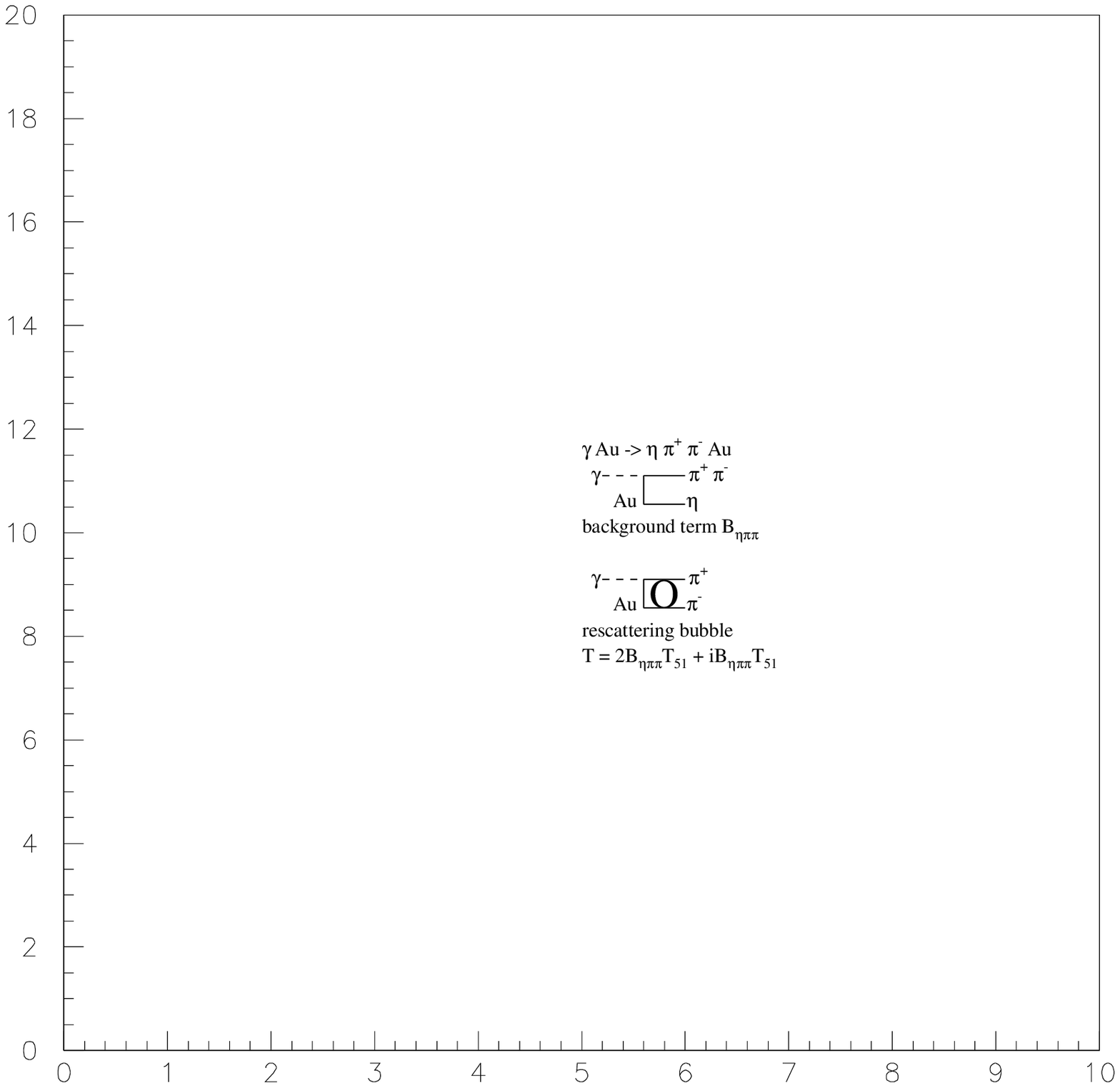}}
\end{center}
\vspace{2pt}
\caption{Beyond re-scattering through $\pi$ $\pi$ loops we focus on two terms
 one being a direct formation of $\eta$ $\pi$ $\pi$ from the photon and 
the other being the re-scattering through the S-matrix of $\eta$ $\pi$ $ \pi$ 
to $\pi$ $\pi$. The loop or bubble has a real and a imaginary part. 
This leads to the above equation.}
\label{fig11}
\end{figure}

\subsection{The Higher Mass Region Poles and Unitarity}
  
In the S-matrix in the higher mass region there are 3 poles plus the 
possibility of background terms when we consider $T_{11}$ and $T_{51}$. Unitarity
means that all of these terms must fit together and satisfy the unitary 
constraints. We are able to achieve such an unitary construction using an 
ad hoc step by step construction of first $T_{11}$ and then $T_{51}$. The tail
of the $\rho(760)$ meson extend into the high mass region tailing off to a
small amplitude. We will denote this background as $B_{11}$. The shape of this
background is shown in Figure 15. The poles for $\rho(1330)$ and $\rho(1730)$
have a very small coupling to the $\pi$ $\pi$ channel. For these poles we use
a Breit-Wigner form which has the same pole position as our k-matrix fit. 
Let us denote these forms as $BW_1$($\rho(1330)$) and $BW_{3}$($\rho(1730)$).
We define the quasi factorizable Breit-Wigner form($\tilde {BW_2}$) through
the equation

\begin{equation}
\tilde {BW_2} = T_{11} - B_{11} - C_{111} BW_1 - C_{113} BW_3.
\end{equation}

where $C_{111}$ is a complex constant that makes the $\rho(1330)$ pole residue
the same as the k-matrix fit, and $C_{113}$ is also a  complex constant that 
makes the $\rho(1730)$ pole residue the same. We turn to the $T_{51}$ amplitude
and define a background term $B_{51}$ such that 

\begin{equation}
B_{51} = T_{51} - C_{512}\tilde {BW_2} - C_{511} BW_1 - C_{513} BW_3.
\end{equation}

The above equation has the constraint

\begin{equation}
REAL(T_{51} - C_{512}\tilde {BW_2} - C_{511} BW_1 - C_{513} BW_3) \approx 0.
\end{equation}

The final fit to the high mass photoproduction data is given by

\begin{equation}
T = 2.0 B_{\pi\pi} T_{11} + B_{\pi\pi} (1.0 + i T_{11}) + C_1 T_{51} + C_2 B_{11} + C_3 B_{51} + C_4 BW_1 + C_5 BW_3.+ C_6 \tilde {BW_2}.
\end{equation}

The $B_{\pi\pi}$ is a real constant and $C_1, C_2, C_3, C_4, C_5, C_6$ are 
complex constants. The results of the eight terms are shown in figures 13 
through 19 with the final sum being Figure 20 with the fit shown in Figure 21.
The overall $\chi^2$ for this global fit which has 1649 degrees of freedom is 
1649. In the higher mass region we get an improvement of 20 in $\chi^2$. This
is a better fit than before but not statistical significant. We can compare
with the higher mass amplitude from the S$\rm\ddot o$ding Model fit which we
show in Figure 22.

\section{Summary and Discussion}

We have perform a global fit to the photoproduction data using a generalized 
S$\rm\ddot o$ding Model\cite{soding} plus $e^+$ $e^-$ scattering to $\pi^+$ 
$\pi^-$\cite{eepipi} and $\pi^+$ $\pi^-$ $\pi^0$\cite{eepipipi}. To this fit 
we also use p-wave partial wave analysis of $\pi^+$ $\pi^-$ to $\pi^+$ 
$\pi^-$\cite{pipipipi}. The 4 channel k-matrix is used for low mass region and 
the 5 channel for high mass region as set up in section 2. We fit directly
to the $T_{11}$ Argand plot of Ref.\cite{pipipipi}. As part of the global fit
we also fit Ref.\cite{eepipi} and Ref.\cite{eepipipi} using equation 10 and
11. 

\begin{figure}
\begin{center}
\mbox{
   \epsfysize 8.0in
   \epsfbox{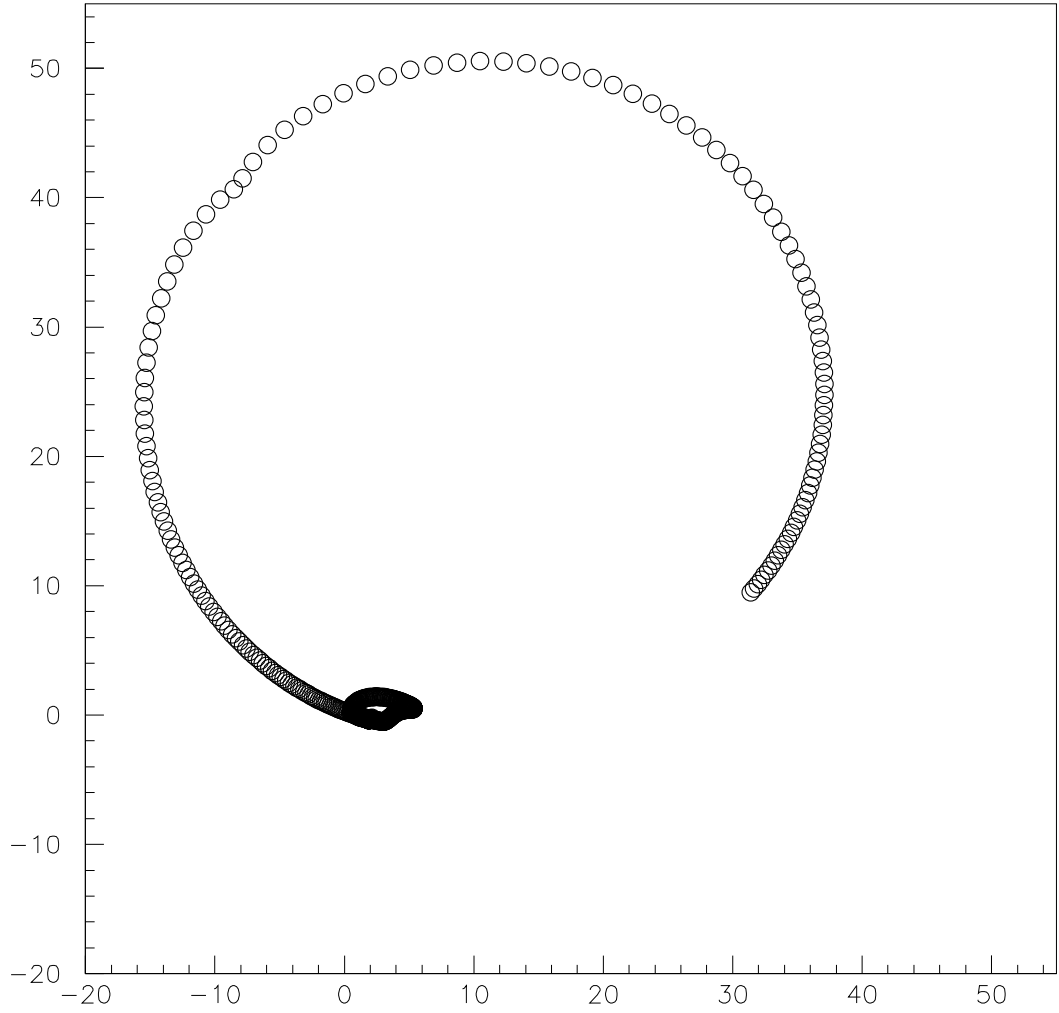}}
\end{center}
\vspace{2pt}
\caption{Equation 14 has two terms one being a direct formation of $\pi^+$ 
$\pi^-$ from the photon and the other being the re-scattering through the 
S-matrix of $\pi$ $ \pi$ p-wave scattering. This loop or bubble has a real 
and a imaginary part. In the fit of equation 20 to the STAR high precision 
Au + Au ultra-peripheral coherent photoproduction data at $\sqrt{s_{NN}} =$
200 GeV(the highest RHIC energy) we obtain the above amplitude for equation 14.}
\label{fig12}
\end{figure}

\begin{figure}
\begin{center}
\mbox{
   \epsfysize 8.0in
   \epsfbox{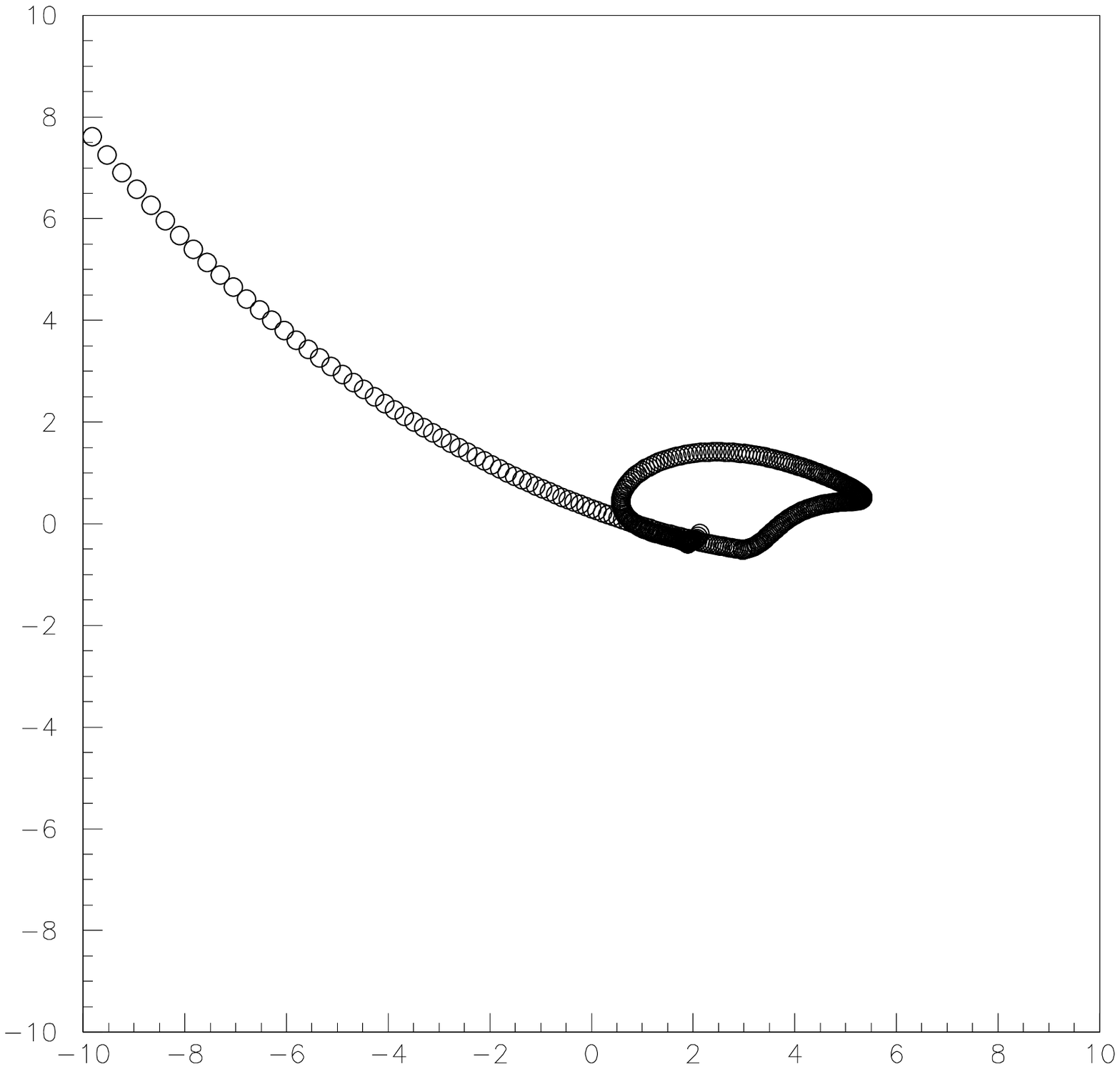}}
\end{center}
\vspace{2pt}
\caption{Equation 14 has two terms one being a direct formation of $\pi^+$ 
$\pi^-$ from the photon and the other being the re-scattering through the 
S-matrix of $\pi$ $ \pi$ p-wave scattering. This loop or bubble has a real 
and a imaginary part. In the fit of equation 20 to the STAR high precision 
Au + Au ultra-peripheral coherent photoproduction data at $\sqrt{s_{NN}} =$ 
200 GeV(the highest RHIC energy) we obtain the above equation 14 amplitude 
for the higher mass region 1.0 GeV to 1.9 GeV.}
\label{fig13}
\end{figure}

\begin{figure}
\begin{center}
\mbox{
   \epsfysize 8.0in
   \epsfbox{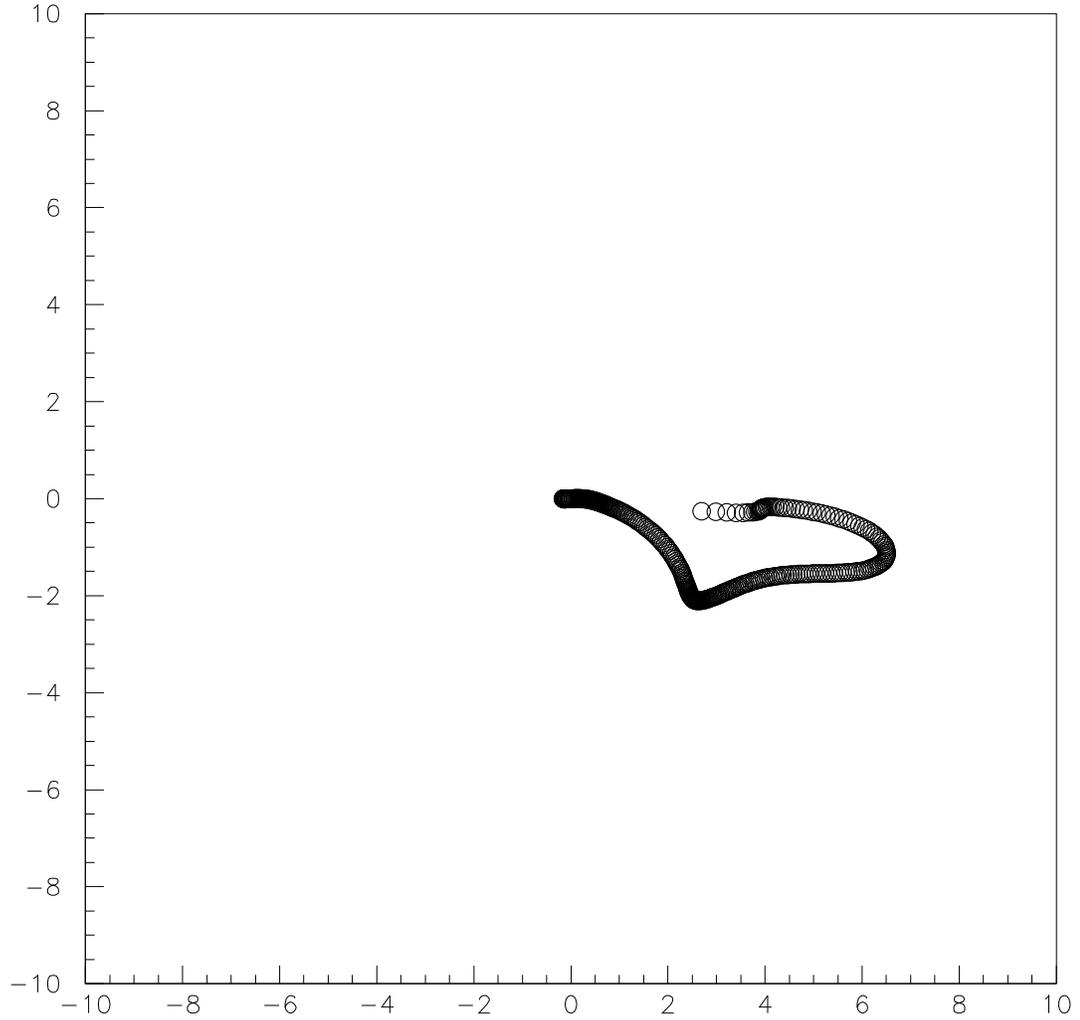}}
\end{center}
\caption{Equation 16 has two terms coming from a loop or bubble re-scattering 
through the S-matrix of $\eta$ $\pi$ $ \pi$ to $\pi$ $\pi$. The loop or bubble 
has a real and a imaginary part. In the fit of equation 20 to the STAR high 
precision Au + Au ultra-peripheral coherent photoproduction data at 
$\sqrt{s_{NN}} =$ 200 GeV(the highest RHIC energy) we obtain for the third 
term(equation 16) the above amplitude for the higher mass region 1.0 GeV to 
1.9 GeV.}
\label{fig14}
\end{figure}

\begin{figure}
\begin{center}
\mbox{
   \epsfysize 8.0in
   \epsfbox{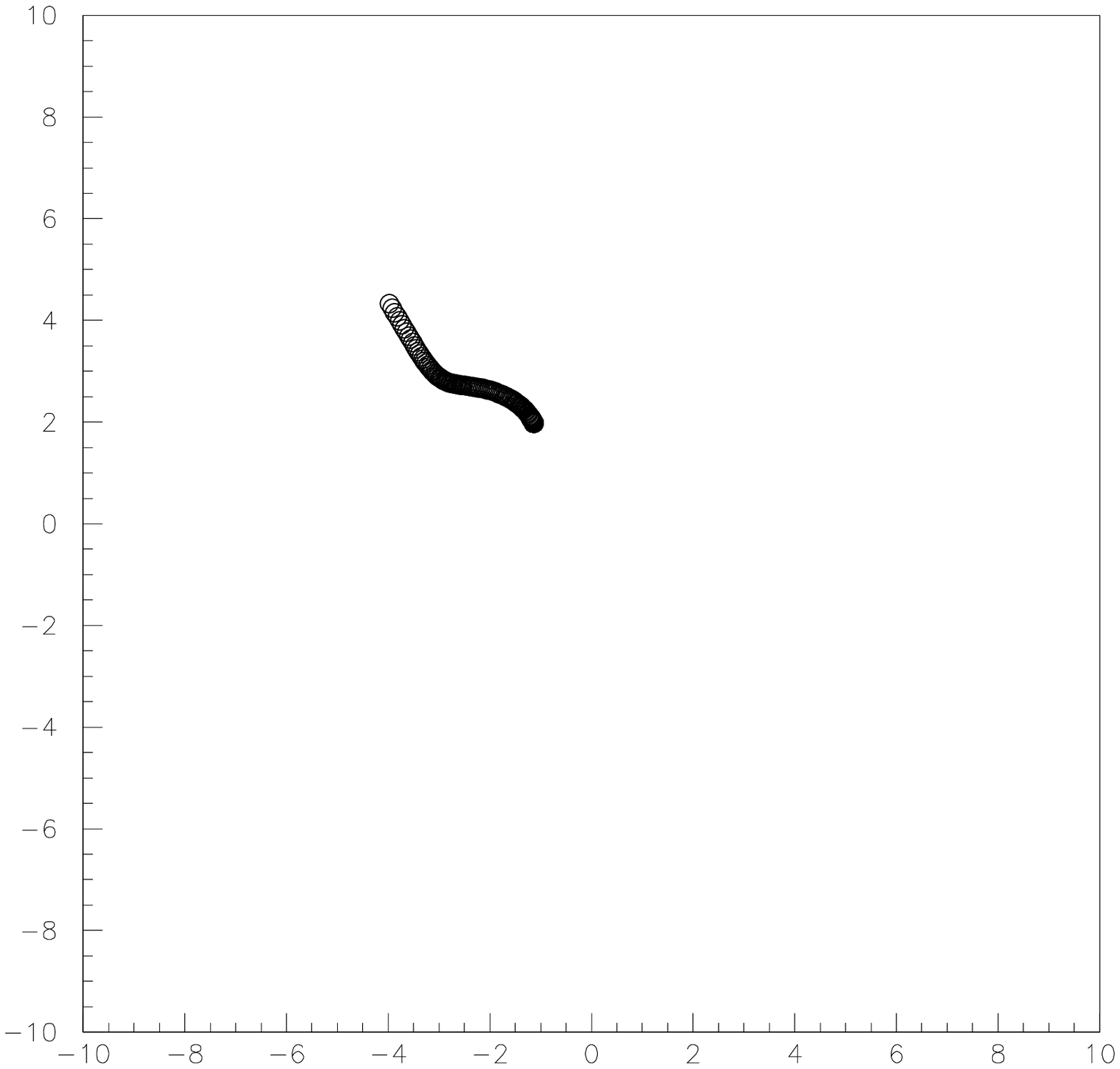}}
\end{center}
\caption{In the S-matrix $\pi$ $ \pi$ to $\pi$ $\pi$ there is a smooth 
background term which is used in equation 20(the fourth term). In the fit of 
equation 20 to the STAR high precision Au + Au ultra-peripheral coherent 
photoproduction data at $\sqrt{s_{NN}} =$ 200 GeV(the highest RHIC energy) 
we obtain for the fourth term the above amplitude for the higher mass region 
1.0 GeV to 1.9 GeV.}
\label{fig15}
\end{figure}

\begin{figure}
\begin{center}
\mbox{
   \epsfysize 8.0in
   \epsfbox{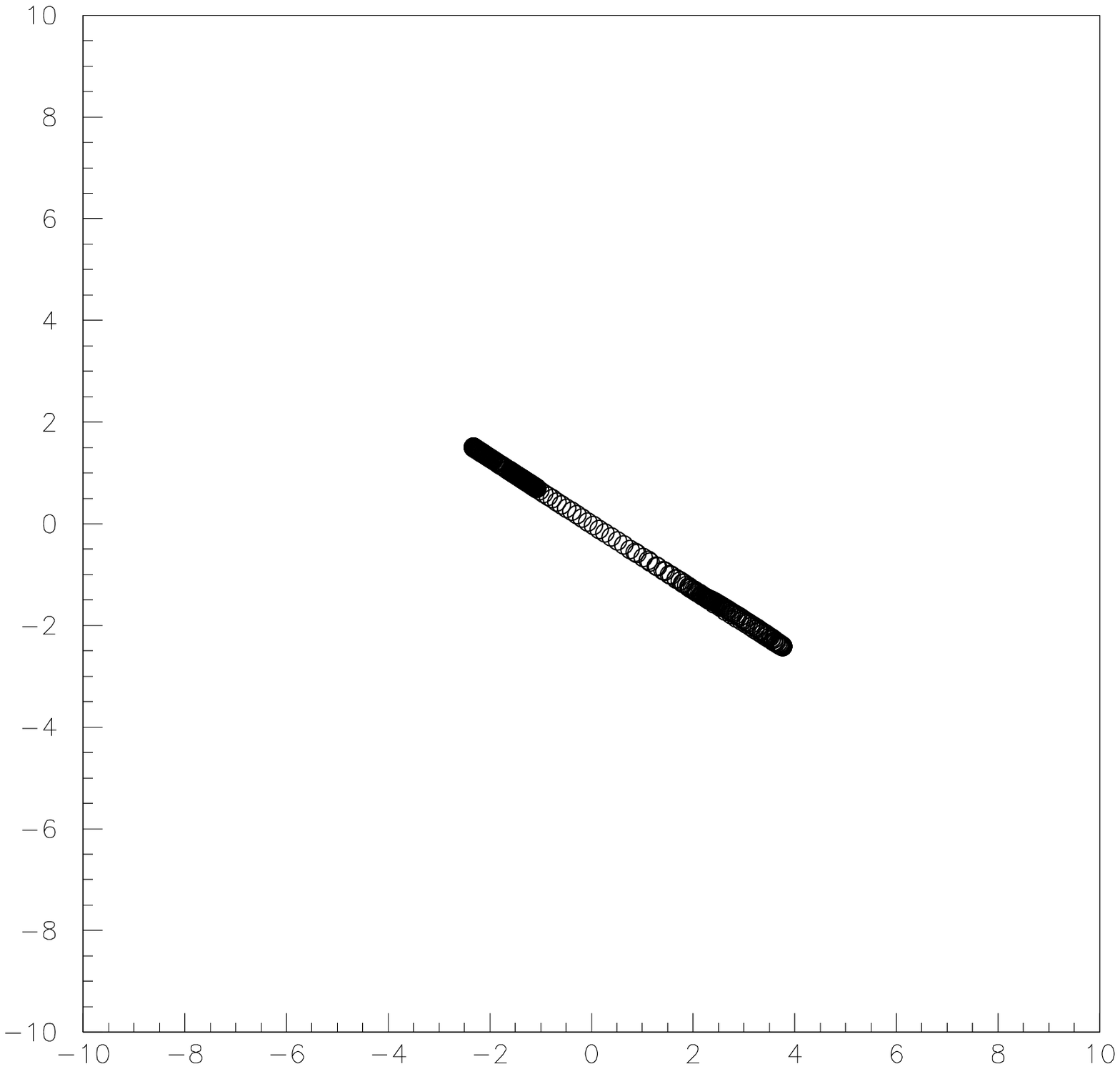}}
\end{center}
\caption{In the S-matrix $\eta$ $\pi$ $ \pi$ to $\pi$ $\pi$ there is a smooth 
background term which is used in equation 20(the fifth term). In the fit of 
equation 20 to the STAR high precision Au + Au ultra-peripheral coherent 
photoproduction data at $\sqrt{s_{NN}} =$ 200 GeV(the highest RHIC energy) we 
obtain for the fifth term the above amplitude for the higher mass region 1.0 
GeV to 1.9 GeV.}
\label{fig16}
\end{figure}

\begin{figure}
\begin{center}
\mbox{
   \epsfysize 8.0in
   \epsfbox{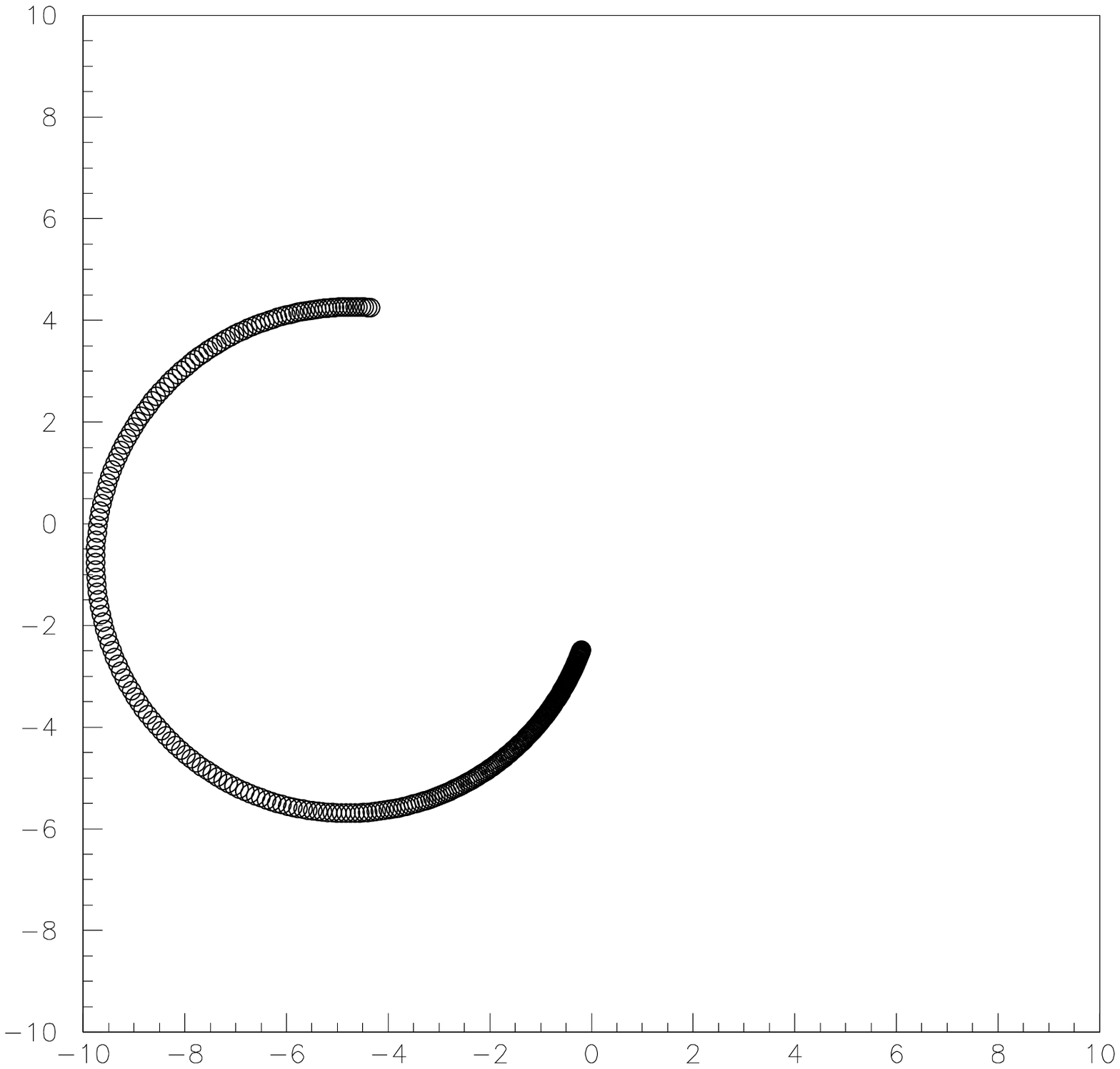}}
\end{center}
\caption{The Breit-Wigner pole($BW_1$($\rho(1330)$)) for the $\rho(1330)$
is used in equation 20(the sixth term). In the fit of equation 20 to the STAR 
high precision Au + Au ultra-peripheral coherent photoproduction data at 
$\sqrt{s_{NN}} =$ 200 GeV(the highest RHIC energy) we obtain for the sixth 
term the above amplitude for the higher mass region 1.0 GeV to 1.9 GeV.}
\label{fig17}
\end{figure}

\begin{figure}
\begin{center}
\mbox{
   \epsfysize 8.0in
   \epsfbox{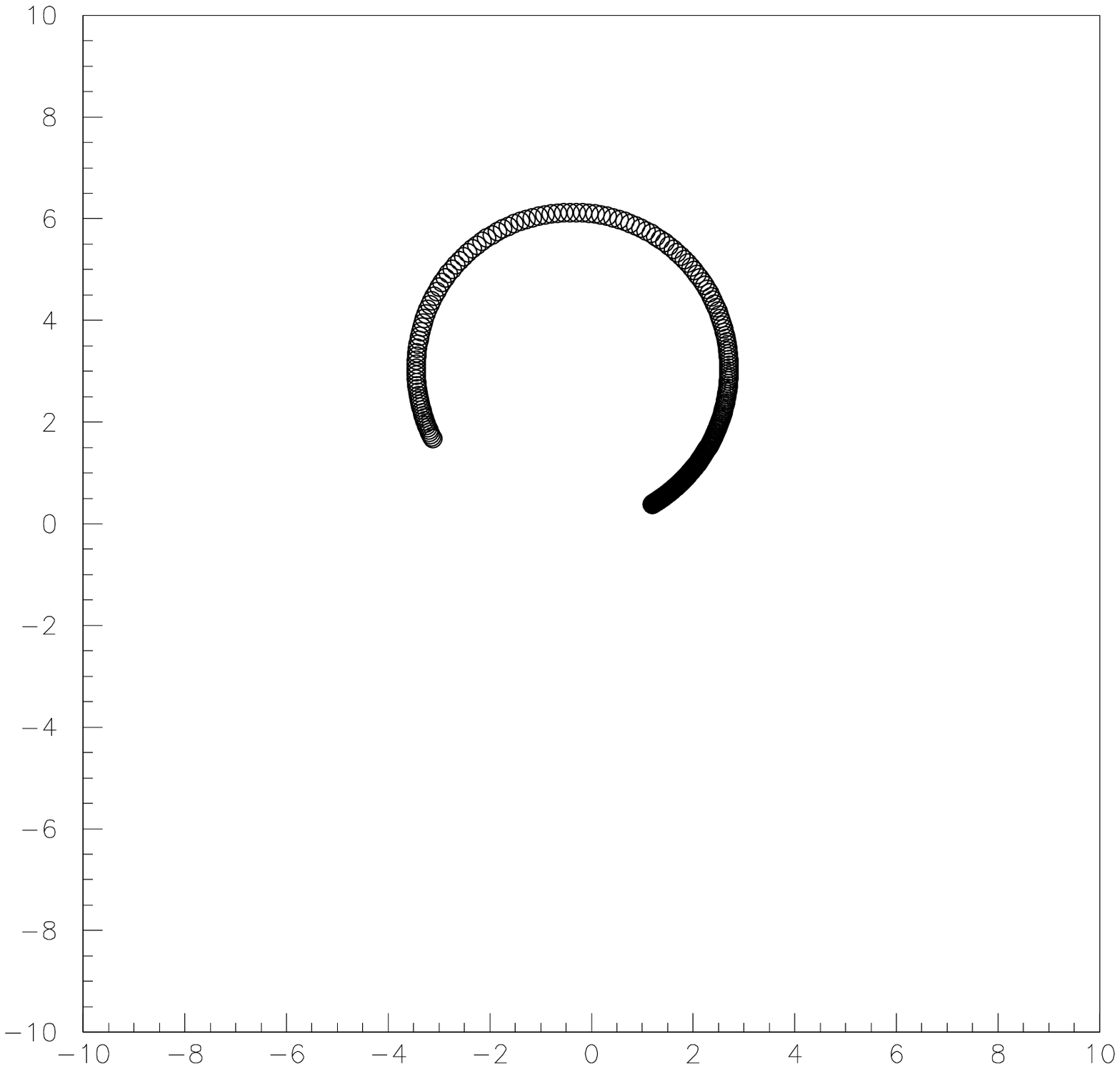}}
\end{center}
\caption{The Breit-Wigner pole($BW_{3}$($\rho(1730)$)) for the $\rho(1730)$
is used in equation 20(the seventh term). In the fit of equation 20 to the STAR 
high precision Au + Au ultra-peripheral coherent photoproduction data at 
$\sqrt{s_{NN}} =$ 200 GeV(the highest RHIC energy) we obtain for the seventh 
term the above amplitude for the higher mass region 1.0 GeV to 1.9 GeV.}
\label{fig18}
\end{figure}

\begin{figure}
\begin{center}
\mbox{
   \epsfysize 8.0in
   \epsfbox{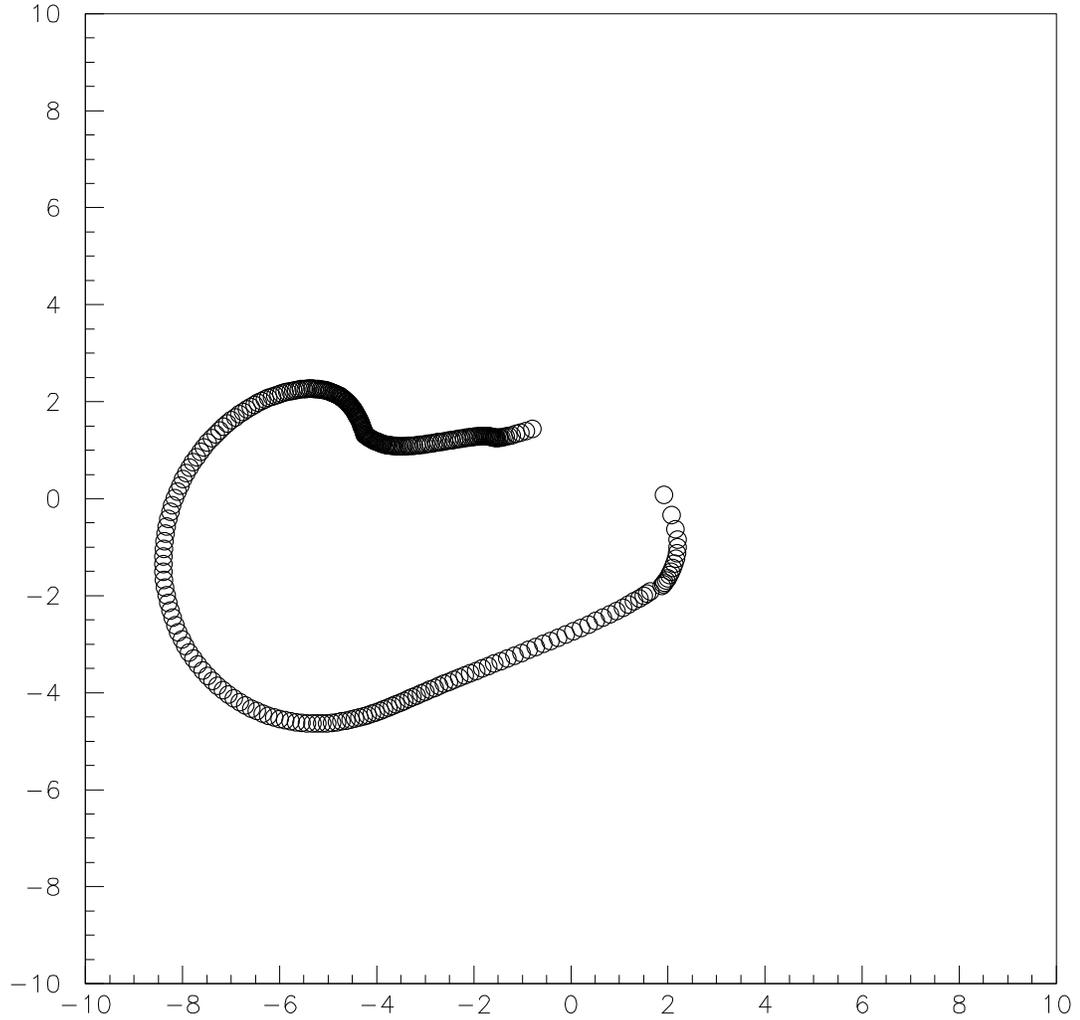}}
\end{center}
\caption{The unitary modified Breit-Wigner pole($\tilde {BW_2}\rho(1630)$)) 
for the $\rho(1630)$ is used in equation 20(the eighth term). In the fit of 
equation 20 to the STAR high precision Au + Au ultra-peripheral coherent 
photoproduction data at $\sqrt{s_{NN}} =$ 200 GeV(the highest RHIC energy) we 
obtain for the eighth term the above amplitude for the higher mass region 
1.0 GeV to 1.9 GeV.}
\label{fig19}
\end{figure}

\begin{figure}
\begin{center}
\mbox{
   \epsfysize 8.0in
   \epsfbox{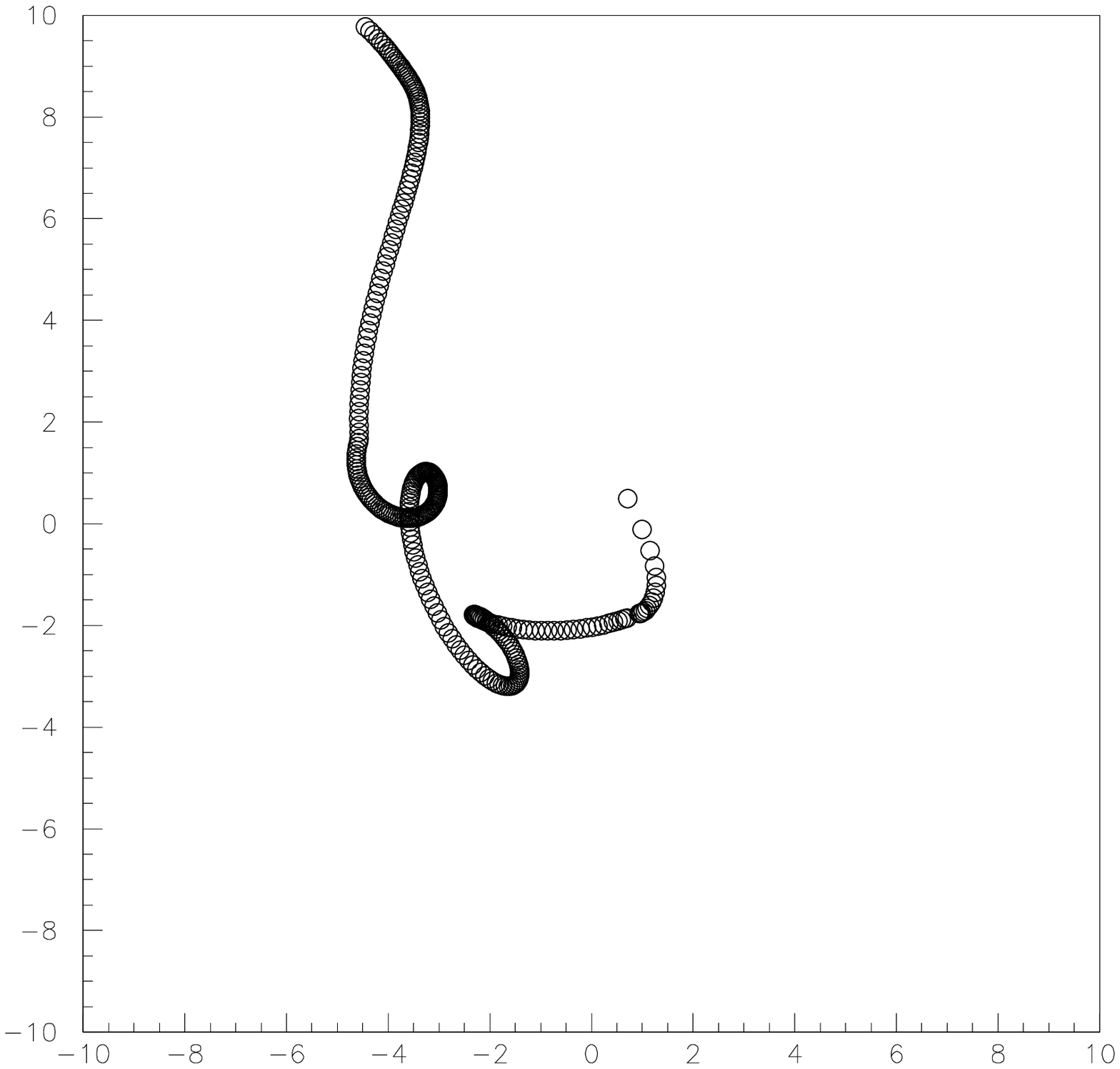}}
\end{center}
\caption{In the fit of equation 20 to the STAR high precision Au + Au 
ultra-peripheral coherent photoproduction data at $\sqrt{s_{NN}} =$ 
200 GeV(the highest RHIC energy) we obtain the above amplitude for the higher 
mass region 1.0 GeV to 1.9 GeV.}
\label{fig20}
\end{figure}

\begin{figure}
\begin{center}
\mbox{
   \epsfysize 8.0in
   \epsfbox{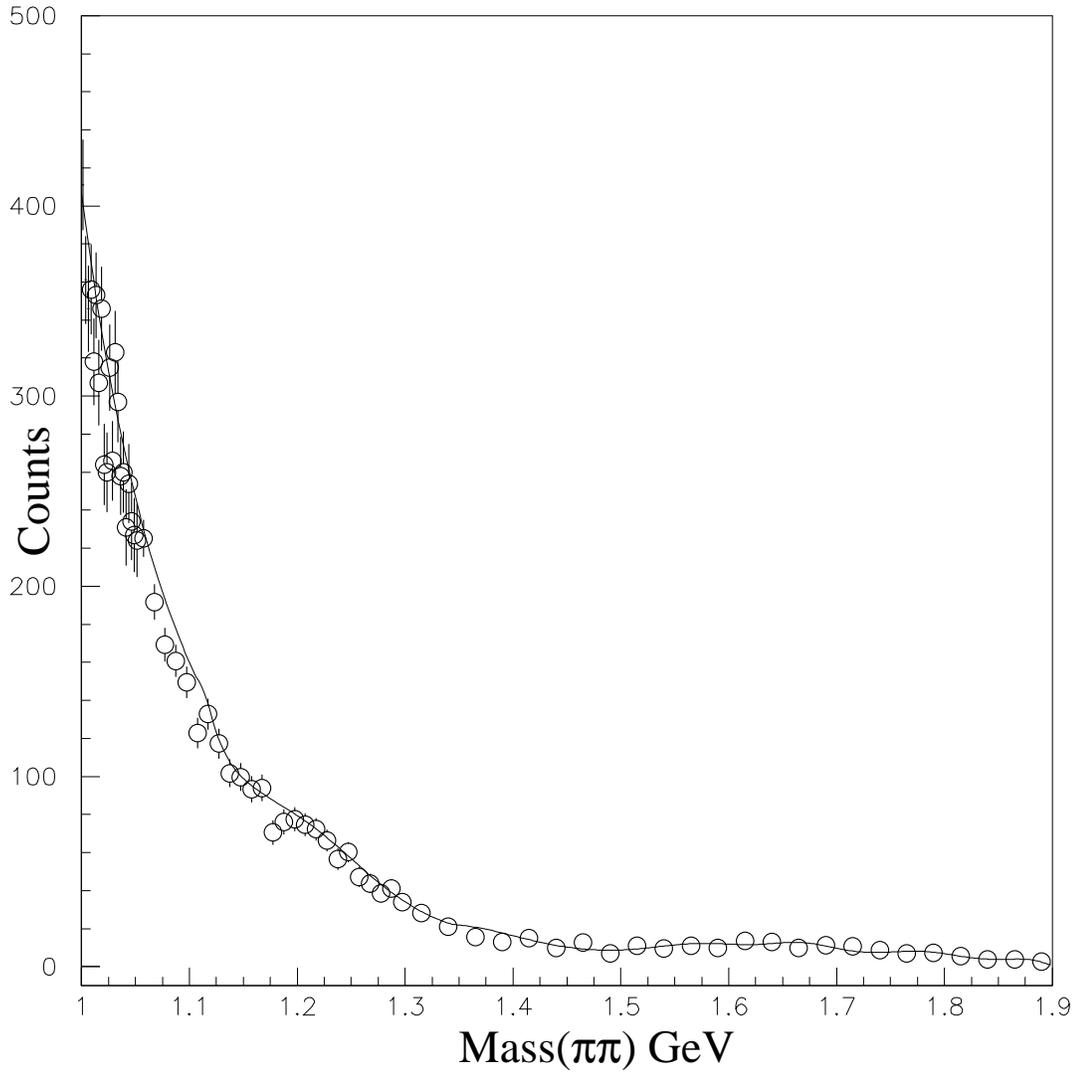}}
\end{center}
\caption{ The fit to the STAR high precision Au + Au ultra-peripheral 
coherent photoproduction data at $\sqrt{s_{NN}} =$ 200 GeV(the highest RHIC 
energy) for the higher mass region 1.0 GeV to 1.9 GeV with the amplitude 
coming from equation 20.}
\label{fig21}
\end{figure}

\begin{figure}
\begin{center}
\mbox{
   \epsfysize 8.0in
   \epsfbox{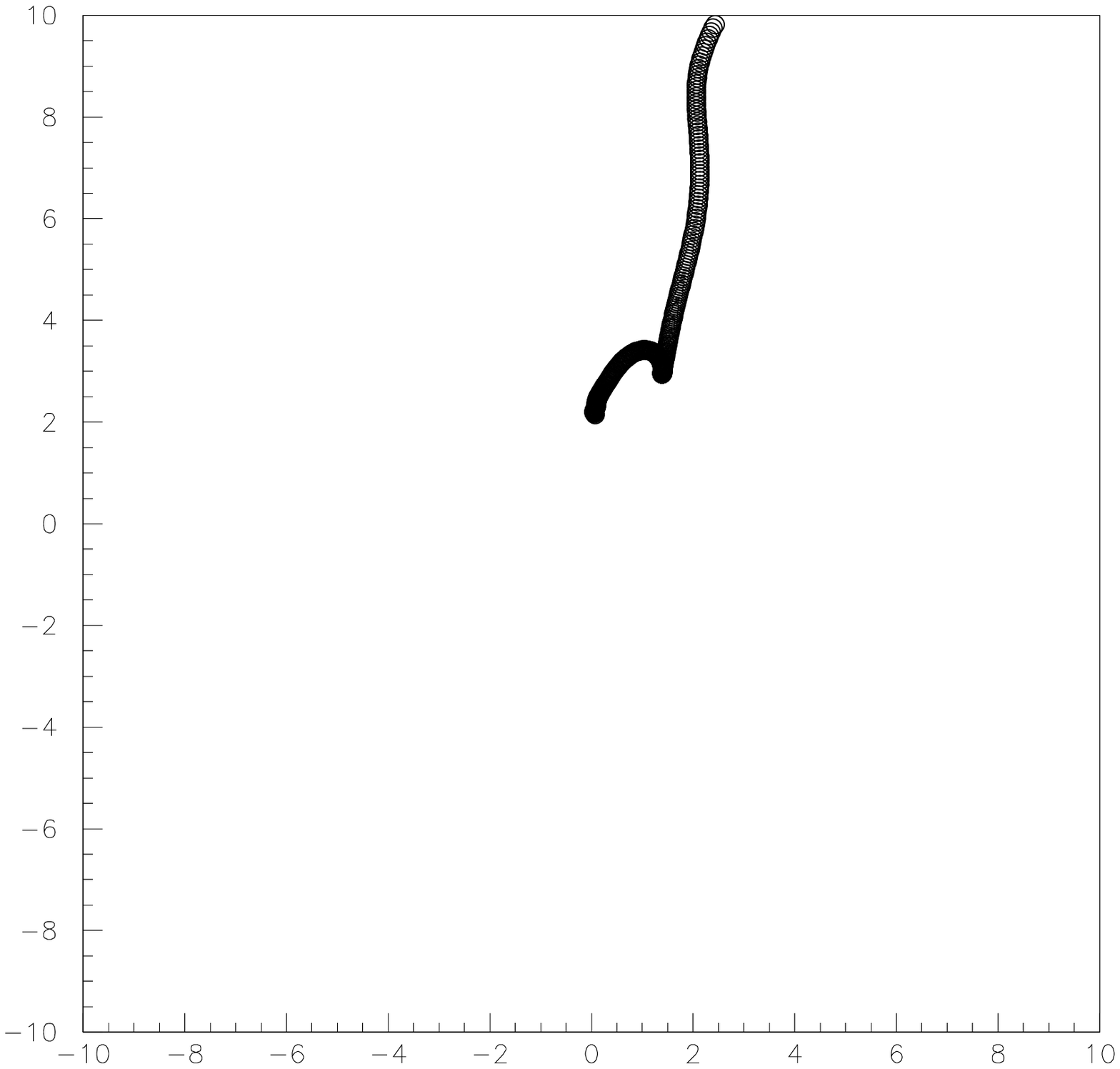}}
\end{center}
\caption{In the fit of the S$\rm\ddot o$ding Model to the STAR high precision 
Au + Au ultra-peripheral coherent photoproduction data at $\sqrt{s_{NN}} =$ 
200 GeV(the highest RHIC energy) we obtain the above amplitude for the higher 
mass region 1.0 GeV to 1.9 GeV.}
\label{fig22}
\end{figure}

\clearpage

The coherent photoproduction data are fitted using a generalized 
S$\rm\ddot o$ding Model\cite{soding} which is outlined in Figure 1. We see in 
the figure that there are two terms. One being a S$\rm\ddot o$ding background 
term of pion pairs coming directly out of the vacuum plus a second being the 
photon coupling directly to the $1^{--}$ scattering matrix. The universal 
lepton pair coupling through the photon gives use $T_{31}$ as the direct 
coupling. We use the form for the S$\rm\ddot o$ding background term given by 
equation 12. The directly coupled term of the photon to the S-matrix is given 
by equation 13. The results of the global fit is shown in Figures 2 through 6. 
The overall $\chi^2$ for the global fit which has 1669 degrees of freedom is 
1926. The 1 $\sigma$ error on such a fit is a $\Delta \chi^2$ of 58. This 
implies that this global fit is 4.8$\sigma$ away from a 1669 $\chi^2$ fit.

We have achieve a global fit to the data sets which gives an analytic form 
for the S-matrix. Using this analytic form we search for poles in the 
$1^{--}$ S-matrix. In this search we find 5 poles. The first pole is the 
well know $\rho (760)$, while its isospin partner $\omega (780)$ is the 
second pole. In the high mass region one expects there should be two radial 
excitations of the $\rho (760)$. These excitation masses should be at around 
1300 and 1800 MeV. One should also expect there should be a d-wave 
$q$ $\bar q$ state at around 1600 MeV. We see the poles in the high mass are 
$\rho(1330)$, $\rho(1630)$ and $\rho(1730)$ which is consistent with this
picture.

The S$\rm\ddot o$ding Model has a built-in term of interference between the
direct production term and the background term. This interference term is
equivalent to a re-scattering as shown in Figure 7. The re-scattering term 
is a loop of pions coming from the background and re-interacting with the
S-matrix of $\pi$ $ \pi$ p-wave scattering. The loop or bubble has a real and a 
imaginary part. The imaginary part of the loop is equal to 
$B_{\pi\pi}$(equation 12) times S-matrix $\pi$ $ \pi$ p-wave scattering. The 
real part of the loop is equal to $\alpha B_{\pi\pi}$ times S-matrix $\pi$ $\pi$
scattering. The value of $\alpha$ is determined to be 2.0 in Ref.\cite{loop} 
from photo production data. Figure 7 and equation 14 gives the equation for 
the background plus re-scattering. To equation 14 we must add the direct 
production of the $\rho (760)$ pole and the $\omega (780)$ pole in the lower 
mass region. When this is done we break the constraint of the 
S$\rm\ddot o$ding Model that the $\omega (780)$ into $\pi$ $\pi$ has to be the 
same as $e^+$ $e^-$ scattering to $\pi^+$ $\pi^-$\cite{eepipi}. 
With this added freedom the global fit $\chi^2$ improve by 257 which a 
4.8 $\sigma$ improvement(see Figure 8). This photo production amplitude into 
$\pi$ $\pi$ is shown in Figure 9, while the S$\rm\ddot o$ding Model 
$\pi$ $\pi$ amplitude is shown in Figure 10. Even though we have a 
quantitative difference these amplitudes are qualitatively the same.

In the higher mass region a new background becomes possible. In our model for
the $1^{--}$ system we have a large coupling of the $\pi$ $\pi$ channel to the
$\eta$ $\pi$ $\pi$ channel. Thus if the photon directly produces the $\eta$ 
$\pi$ $\pi$ channel, then through a $\eta$ $\pi$ $\pi$ loop one can form
the $\pi$ $\pi$ channel by the cross term of the S-matrix of $\eta$ $\pi$ $ 
\pi$ to $\pi$ $ \pi$(see Figure 11). Again the loop or bubble has a real and 
a imaginary part. The imaginary part of the loop is equal to 
$B_{\eta\pi\pi}$(equation 15) times T-matrix $\eta$ $\pi$ $\pi$ to $\pi$ $\pi$
scattering. The real part of the loop is equal to 2.0 $B_{\eta\pi\pi}$ times 
T-matrix $\eta$ $\pi$ $\pi$ to $\pi$ $\pi$. Figure 11 and equation 16 gives 
the equation for the re-scattering.

We then perform a new global fit to the photoproduction data plus $e^+$ $e^-$ 
scattering to $\pi^+$ $\pi^-$\cite{eepipi} and $\pi^+$ $\pi^-$ 
$\pi^0$\cite{eepipipi} plus the p-wave partial wave analysis of 
$\pi^+$ $\pi^-$ to $\pi^+$ $\pi^-$. As part of this fit we use equation 14 and 
equation 16. Since the photoproduction data is the square of the amplitude 
there is an over all phase that is not determined. Let us choose equation 14
to determine this phase. The result of this fit gives a result for equation 14 
which is displayed in Figure 12. The background rises quickly at threshold and
then falls off with mass. The higher mass region is the focus of Figure 13.
We see that the background term equation 14 for the pion loop is mainly a
real function. Equation 16 which is the $\eta$ $\pi$ $\pi$ loop is also mainly
real and of the same magnitude at these higher masses. This background 
also rises quickly at threshold and then falls off with mass tailing to 
zero(see Figure 14).

In the S-matrix in the higher mass region there are 3 poles plus the 
possibility of background terms when we consider $T_{11}$ and $T_{51}$. Unitarity
means that all of these terms must fit together and satisfy the unitary 
constraints. We are able to achieve such an unitary construction using an 
ad hoc step by step construction of first $T_{11}$ and then $T_{51}$. The tail
of the $\rho(760)$ meson extend into the high mass region tailing off to a
small amplitude. We will denote this background as $B_{11}$. The shape of this
background is shown in Figure 15. The poles for $\rho(1330)$ and $\rho(1730)$
have a very small coupling to the $\pi$ $\pi$ channel. For these poles we use
a Breit-Wigner form which has the same pole position as our k-matrix fit. 
Let us denote these forms as $BW_1$($\rho(1330)$) and $BW_{3}$($\rho(1730)$).
We define the quasi factorizable Breit-Wigner form($\tilde {BW_2}$) through
the equation 17. We then turn to the $T_{51}$ amplitude and define a background 
term $B_{51}$ in equation 18 with the constraint of equation 19.
For the final fit to the high mass photoproduction data we use equation 20
which has eight terms((1and2)$\pi$ $\pi$loops; (3)$\eta$ $\pi$$\pi$loops; 
(4)$T_{11}$background; (5)$T_{51}$background; (6)$\rho(1330)$; (7)$\rho(1730)$:
(8)$\rho(1630)$). The results of the eight terms are shown in figures 13 
through 19 with the final sum being Figure 20 with the fit shown in Figure 21.
The overall $\chi^2$ for this global fit which has 1649 degrees of freedom is 
1649. In the higher mass region we get an improvement of 20 in $\chi^2$. This
is a better fit than before but not statistical significant. We can compare
with the higher mass amplitude from the S$\rm\ddot o$ding Model fit which we
show in Figure 22.

\section{Acknowledgments}

This research was supported by the U.S. Department of Energy under Contract No.
DE-AC02-98CH10886.

\section{Appendix}

The $T_{11}$ elastic scattering amplitude is a complex amplitude
described by two real numbers one bounded between 0.0 and 1.0($\eta_1$) 
and another is in units of angles($\delta_1$). The form of the amplitude is
\begin{equation}
T_{11} = \eta_1 sin\delta_1e^{i \delta_1}
\end{equation}
We note that $\eta_1$ and $\delta_1$ depends on the value of $M_{\pi\pi}$,
and one could also use real($T_{11}$) and imag($T_{11}$).   

In order to make a well controlled k-matrix calculation let us assume that above
a $M_{\pi\pi}$ greater than 1.1 Gev/c there are two important channels 1
$\pi^+$ $\pi^-$ and 2 $\eta$ $\pi^+$ $\pi^-$. Thus we need a 2X2 k-matrix
which we will assume is factorizable with two factors $k_1$ and $k_2$ such that

\begin{equation}
K_{11} = k_1k_1,
   K_{12} = k_1k_2,
   K_{21} = k_1k_2,
   K_{22} = k_2k_2.
\end{equation}

The $T_{11}$ elastic scattering amplitude is equal to

\begin{equation}
T_{11} = \frac{k_1^2}{1.0 - i(k_1^2 + k_2^2)}.
\end{equation}

let us define  

\begin{equation}
\alpha = (k_1^2 + k_2^2),
\end{equation}

and rewriting the above equation we have

\begin{equation}
T_{11} = \frac{k_1^2}{1.0 - i\alpha}.
\end{equation}

We multiply numerator and denominator by the complex conjugate of denominator
of the above equation we obtain

\begin{equation}
T_{11} = \frac{k_1^2 + i\alpha k_1^2}{1.0 + \alpha^2}.
\end{equation}

We see that

\begin{equation}
\frac{T_{11}}{k_1^2} = \frac{1.0 + i\alpha}{1.0 + \alpha^2}.
\end{equation}

Thus the real part

\begin{equation}
\frac{Real(T_{11})}{k_1^2} = \frac{1.0}{1.0 + \alpha^2},
\end{equation}

and the imaginary part

\begin{equation}
\frac{Imag(T_{11})}{k_1^2} = \frac{\alpha}{1.0 + \alpha^2}.
\end{equation}

Dividing equation 29 by 28, we obtain

\begin{equation}
\alpha = \frac{Imag(T_{11})}{Real(T_{11}}.
\end{equation}

\begin{equation}
k_1^2 = (Real(T_{11})(1.0 + \alpha^2).
\end{equation}

\begin{equation}
k_2^2 = \alpha - k_1^2.
\end{equation}

We can redefine the p-wave $\pi$ $\pi$\cite{pipipipi} phase shift into two
real functions between 1.1 Gev to 1.9 Gev. If define a new variable
x such that

\begin{equation}
x = 2.5M_{\pi\pi} - 3.75.
\end{equation}

Over the $M_{\pi\pi}$ range 1.1 Gev to 1.9 Gev,  x varies from -1.0 to 1.0.
Therefore we can expand the two real functions $k_1$ and $k_2$ in terms of
Legendre polynomials. We project out these polynomials up to $9^{th}$ order
and use these values to start the global fits described in the text.


\begin{thebibliography}{99}
\bibitem{photo} L.~Adamczyk {\it et al.}, Phys. Rev. C 96 (2017) 054904.
\bibitem{eepipi} M.N.~Achasov {\it et al.}, arXiv:0605013[hep-ex].
\bibitem{eepipipi} M.N.~Achasov {\it et al.}, arXiv:0305049[hep-ex].
\bibitem{pipipipi} B.~Hyams, C.~Jones, P.~Weilhammer, Nucl. Phys. B 96 (1973)
134-162.
\bibitem{soding} P.~S$\rm \ddot o$ding, Phys. Lett. 19 (1966) 702.
\bibitem{loop} R.S.~Longacre, arXiv:1306.2908[Nucl-th].
\end{thebibliography}
\end{document}